\documentclass[aps,prl,reprint,showpacs,superscriptaddress,amsmath,amssymb]{revtex4-1}

\usepackage{graphicx}
\usepackage{dcolumn}
\usepackage{bm}
\usepackage[colorlinks=true,linkcolor=blue]{hyperref}
\usepackage[outdir=./]{epstopdf}
\usepackage{comment}
\usepackage{color}
\usepackage{xcolor}
\usepackage{bm}
\usepackage{physics}
\usepackage{bbm}
\newcolumntype{L}{@{}>{\kern\tabcolsep}l<{\kern\tabcolsep}}

\begin{document}

\title{Exciton-phonon coupling in the UV absorption and emission spectra of bulk hexagonal boron nitride}

\author{Fulvio Paleari}
\affiliation{Physics and Materials Science Research Unit, University of Luxembourg, 162a avenue de la Fa\"iencerie, L-1511 Luxembourg, Luxembourg}
\author{Henrique P. C. Miranda}
\affiliation{Physics and Materials Science Research Unit, University of Luxembourg, 162a avenue de la Fa\"iencerie, L-1511 Luxembourg, Luxembourg}
\affiliation{Institute of Condensed Matter and Nanosciences, Universit\'{e} catholique de Louvain, Chemin des \'etoiles 8, bte L7.03.01, 1348, Louvain-la-Neuve, Belgium}
\author{Alejandro Molina-S\'{a}nchez}
\affiliation{Institute of Materials Science (ICMUV), University of Valencia, Catedr\'{a}tico Beltr\'{a}n 2, E-46980 Valencia, Spain}
\author{Ludger Wirtz}
\affiliation{Physics and Materials Science Research Unit, University of Luxembourg, 162a avenue de la Fa\"iencerie, L-1511 Luxembourg, Luxembourg}

\date{\today}

\begin{abstract}
We present an \textit{ab initio} method to calculate phonon-assisted absorption and emission spectra in the presence of strong excitonic effects. We apply the method to bulk hexagonal BN which has an indirect band gap and is known for its strong luminescence in the UV range. We first analyse the excitons at the wave vector $\overline{q}$ of the indirect gap. The coupling of these excitons with the various phonon modes at $\overline{q}$ is expressed in terms of a product of the mean square displacement of the atoms and the second derivative of the optical response function with respect to atomic displacement along the phonon eigenvectors. The derivatives are calculated numerically with a finite difference scheme in a supercell commensurate with $\overline{q}$. 
We use detailed balance arguments to obtain the intensity ratio between emission and absorption processes. Our results explain recent luminescence experiments and reveal the exciton-phonon coupling channels responsible for the emission lines.
\end{abstract}

\maketitle

\begin{figure}[t]
\includegraphics[width=0.9\columnwidth]{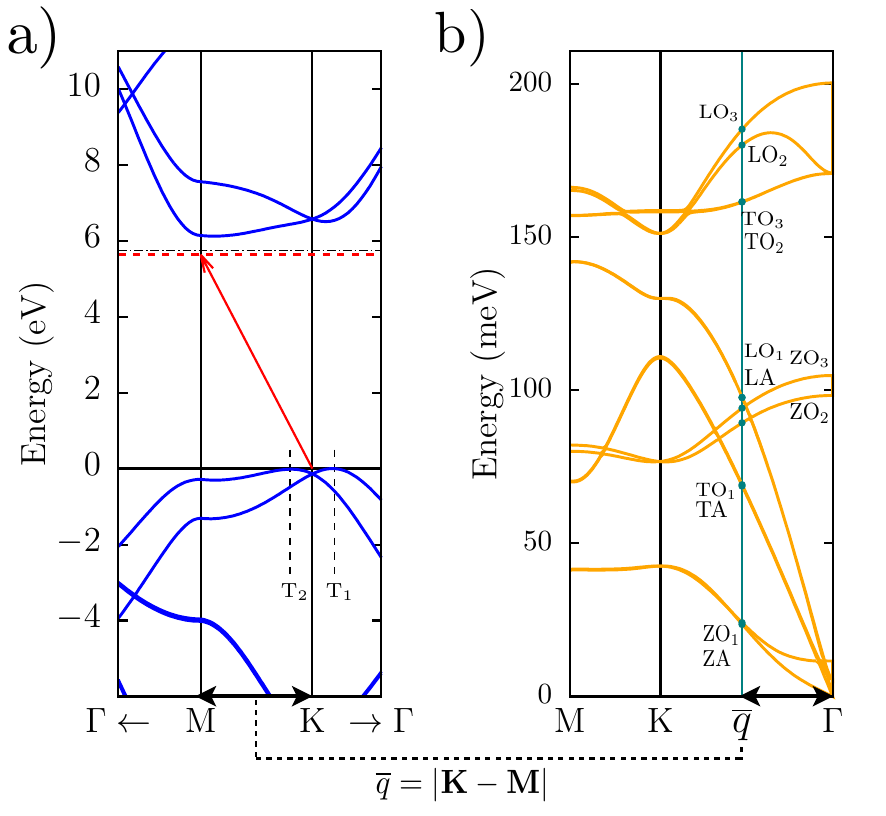}
\caption{Finite-$q$ calculations in hexagonal boron nitride. (a) GW$_0$ quasiparticle band structure of bulk hBN, displayed in the relevant part of the BZ. The dashed horizontal black line represents the position of the lowest-bound direct exciton, while the dashed red lines correspond to the indirect excitons of momentum $\overline{q}=|\mathrm{K}-\mathrm{M}|$. (b) the phonon dispersion in bulk hBN. The teal vertical line highlights the frequencies of the phonon modes with momentum $\overline{q}$.}
\label{fig:1}
\end{figure}

Hexagonal boron nitride (hBN) is well known for its strong luminescence signal in the UV range.\cite{Watanabe2004,Kubota932,Watanabe2009b}
The fine structure of both emission and absorption spectra has been under heavy discussion until very recently.
First-principles calculations agree upon the existence of an indirect \textit{quasiparticle} band gap from the area around the K point to the M point of the hexagonal Brillouin zone (BZ),\cite{Arnaud2006} with the direct band gap lying $0.5$ eV higher (Fig.~\ref{fig:1} (a)).
However, the layered structure with a quasi 2D confinement of electron-hole pairs within the layer and reduced screening outside leads to the formation of strongly bound excitons \cite{Arnaud2006}.
In experimental absorption spectra\cite{Lauret2005}, a strong peak is seen around $6$ eV, which is interpreted by first-principles calculations using many-body perturbation theory (MBPT) as an exciton with a huge binding energy of $700$ meV.
This \textit{direct} exciton lies below the bottom of the conduction band (black dashed line in Fig.~\ref{fig:1} (a)). 
Therefore, the \textit{optical} gap of hBN has often been considered to be ``direct'', despite experimental observation of a fine structure appearing in both absorption\cite{Watanabe2004} and emission spectra\cite{Jaffrennou2007,Watanabe2009}. Symmetry-breaking effects such as Jahn-Teller distortion in the excited state\cite{Watanabe2009} or the interaction of excitons with point defects\cite{PhysRevB.83.144115} were invoked to explain the fine structure.

Recently, with the help of photoluminescence on high-purity samples, Cassabois \textit{et al.}\cite{CassaboisG.2016} explained the fine structure in terms of recombination of electrons from the conduction band minimum to the valence band maximum, assisted by emission of phonons with the corresponding wave vector $\overline{q}$ (See Fig.~\ref{fig:1}).
Several new publications followed up on this topic, cementing this interpretation\cite{Cassabois2016,exp_data,PhysRevB.94.121405,Schue2018arXiv} notably through the mass-dependence of the position of phonon-assisted emission peaks in isotopically clean samples \cite{vuong:hal-01698228}. 
Yet, the proposed interpretation is on the level of independent electrons and holes, insufficient to account for the strong Coulomb interaction in hBN and thus neglecting excitonic effects. Clearly, the position of the direct exciton is below the conduction band minimum in Fig.~\ref{fig:1}. However, excitonic states with finite wave vector $\overline{q}$ can have an energy below the direct exction (horizontal red-dashed line). The position of these indirect excitons and how they contribute to optical absorption and photoluminescence via emission and absorption of phonons of wave vector $\overline{q}$ are the topic of this letter.

\textit{Ab initio} calculations of indirect absorption are still sparse and, so far, restricted to the independent particle-picture. Notably, indirect absorption was calculated for silicon\cite{PhysRevLett.108.167402,patrick,Marios1,Marios2} where the independent-particle picture is a good approximation.
A unified approach to describe on the same footing lattice-dependent band features (such as band gap renormalization with temperature) and phonon-assisted transitions was proposed\cite{patrick,Marios1,Marios2}. However, it is restricted to the static approximation and it is computationally expensive.
Moreover, a perturbation-theory treatment of the exciton-phonon coupling including dynamical effects was proposed,\cite{Antonius2017} extending previous works,\cite{toyozawa, PhysRev.171.935, Rudin1990, PhysRevLett.101.106405} although its direct test on real materials is, again, computationally demanding.
Here, we present a computational scheme that captures the exciton-phonon coupling in a finite-difference approach and treats luminescence intensities using detailed balance arguments.
This enables us to explain the experimental PL spectra in terms of phonon assisted recombination from finite-momentum excitonic states.

Hexagonal BN is a layered material with a honeycomb lattice in AA' stacking, possessing $D_{6h}$ space group symmetry.
Using experimental lattice parameters\cite{SOLOZHENKO19951}, we first compute the direct excitons using the primitive unit cell.
Starting with the independent-particle band structure at the the DFT level, we obtain the quasiparticle band structure through a semi-self-consistent GW$_0$ scheme, converging iteratively only the eigenvalues entering the Green's function. 
This procedure is used to correct the usual quasiparticle gap underestimation given by the G$_0$W$_0$ approximation\cite{Hybertsen1986} in BN systems.\cite{multiBN}
We obtain a direct quasiparticle gap of $6.46$ eV, and an indirect one of $5.96$ eV (Fig. \ref{fig:1}a).
In the following we will approximate the momentum difference of the indirect band gap as $\overline{q}=|\mathrm{\textbf{K}}-\mathrm{\textbf{M}}|$. A discussion on this approximation with respect to the exact positions T$_1$ and T$_2$\cite{exp-band-bulk} (Fig. \ref{fig:1}a) can be found in the Supplementary Information (SI), Sec.~III.

\begin{figure}[t!]
\includegraphics[width=0.9\columnwidth]{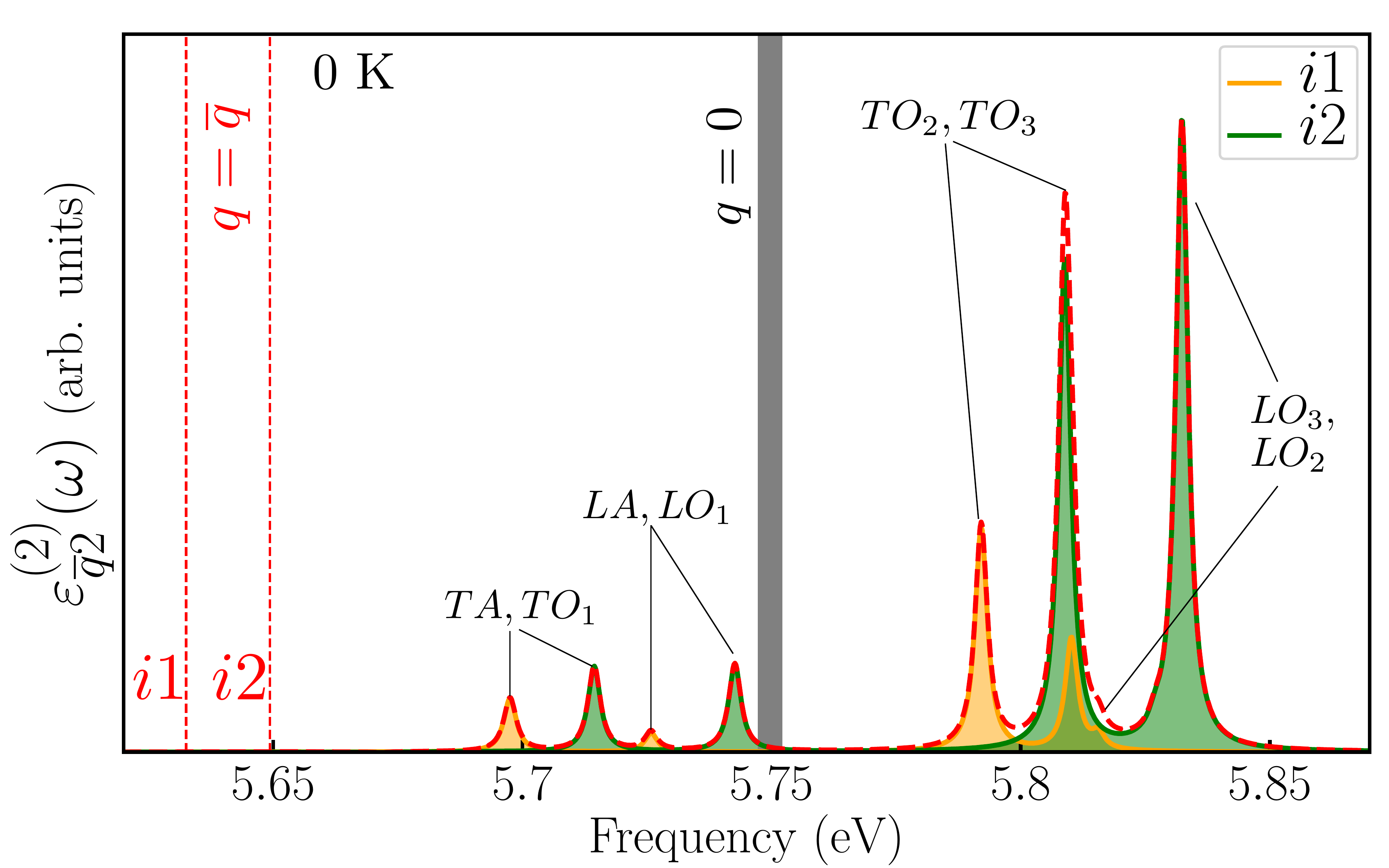}
\caption{Phonon-assisted absorption in hexagonal boron nitride. Imaginary part of the dielectric function including contributions of transitions at $q=\overline{q}$ mediated by a single phonon (dashed red line, $\varepsilon^{(2)}_{\overline{q}2}(\omega)$ in the text). The orange (green) peaks originate from the phonon couplings to $i1$ ($i2$) and the phonon modes responsible for them are labeled. The peak broadenings are set to $1.5$ meV for the temperature of $0$ K. The dashed red vertical lines indicate the positions of the two pairs of A$_1$+B$_1$ excitons at $\overline{q}$ (labeled $i1$ and $i2$). The thick gray vertical line is at the position of the main optically active $q=0$ exciton. \label{fig:2}}
\end{figure}

Solving the Bethe-Salpeter equation (BSE)\cite{Rohlfing2000}, we obtain two doubly degenerate excitonic states at $q=0$ which form a Davydov pair\cite{Davydov,Dawson1975} due to interlayer interaction\cite{multiBN}: The first exciton at $5.70$ eV has $E_{2g}$ symmetry and is thus \textit{dark}. The second exciton at $5.75$ eV (gray vertical line in Fig.~\ref{fig:2}) possesses $E_{1u}$ symmetry (odd under inversion) and is thus \textit{bright}. (More on the group theory can be found in section VI of the SI and in Ref.~\onlinecite{multiBN}. Note that we are only dealing with light polarized parallel to the BN planes which dominates the optical spectra.)

In order to study finite momentum $\overline{q}$ excitons, we perform the same calculations on a special, non-diagonal hBN supercell, containing 12 atoms per layer and chosen such that $\overline{q}$ will be folded onto $\Gamma$ in its new Brillouin zone\cite{lloyd} (see SI).
Two additional excitonic states, $i1$ and $i2$ (dashed red lines in Fig. \ref{fig:2}) at $5.63$ and $5.65$ eV, respectively, appear below the lowest-bound direct exciton (in agreement with recent results\cite{Schue2018arXiv,finiteq} and confirming that not only the quasiparticle gap but also the optical gap of bulk hBN is ``indirect''). These states originate from the splitting, at finite momentum, of the doubly degenerate $E_{2g}$ exciton. 
The $i1$ exciton transforms as the $B_1$ representation of the $C_{2v}$ symmetry group, and $i2$ transforms as $A_1$ (see SI).
These finite-$q$ states are \textit{dark} by themselves but candidates for phonon-assisted absorption and emission. 
The same symmetry considerations can be found in the phonon dispersion (Fig. \ref{fig:1}b): At $\Gamma$, the in-plane phonon modes form Davydov pairs and transform according to the $E_{1u}$ and $E_{2g}$ representation, respectively. Along $\Gamma \mathrm{K}$ these representations reduce to $A_1$ and $B_1$.
According to group theory (see SI), the $B_1$ exciton can couple to $B_1$ phonons and the $A_1$ exciton can couple to $A_1$ phonons if the light polarization $\mathbf{e}$ is parallel to $\overline{q}$. For light polarization perpendicular to $\ \overline{q}$, the $B_1$ exciton couples to $A_1$ phonons and the $A_1$ exciton to $B_1$ phonons. Coupling with out-of-plane modes is forbidden.\cite{VuongZmodes}

Having understood the symmetry constraints for coupling between finite-$q$ excitons and phonons, we now derive a general expression to calculate these couplings and, thus, phonon-assisted optical spectra. We use (i) a static approximation for the exciton formation probability and (ii) restricting the coupling to harmonic phonons with momentum $\overline{q}$. In our calculations, we focus on the energy region close to $i1$ and $i2$. We proceed by considering the complex dielectric function $\varepsilon(\omega)=\varepsilon_1(\omega)+\mathrm{i}\varepsilon_2(\omega)$, which describes the linear response of the system and is closely related to the experimentally measurable absorption coefficient $\alpha(\omega)=\omega \varepsilon_2(\omega)/c n_r[\varepsilon(\omega)]$ ($n_r$ being the refractive index).
A calculation with the atoms clamped at the equilibrium positions yields $\varepsilon^{(0)}_2(\omega)=C\sum_S |T^S|^2 \Im{1/(\hbar\omega -E^S+\mathrm{i} \eta) }$ where $|T^S|^2$ is the formation probability of exciton $S$ and $C$ is a dimensional constant (Sec. I in SI). 
The denominator determines the peak structure of $\varepsilon_2$, with $E^S$ being the exciton energy and $\eta$ the exciton linewidth.
In order to obtain the response due to the indirect excitons, we consider the Taylor expansion of $\varepsilon (\omega)$ up to second-order in the atomic displacements.\cite{Marios2}
This gives a static correction to the equilibrium response $\varepsilon(\omega) = \varepsilon^{(0)}(\omega)+\varepsilon^{\mathrm{st.,}(2)}_{\overline{q}}(\omega) $, in which the term (Sec. V in SI)
\begin{equation}\label{eq:taylor}
\varepsilon^{\mathrm{st.,}(2)}_{\overline{q}}(\omega) =\frac{1}{2}\sum_\lambda\left[ \sum_i^{N_{\overline{q}}} \frac{1}{2}\sum_{j}^2 \frac{\partial^2 \varepsilon^{(0)}_j(\omega)}{\partial R_{\lambda \overline{q}_i}^2}\Bigr|_{\mathrm{eq}} \right]\sigma_{\lambda \overline{q}}^2
\end{equation}
adds the contribution of transitions assisted by a single phonon of momentum $\overline{q}$.
In this expression $j$ represents the polarisation direction of the incoming light, over which we average, and $i$ labels each of the $N_{\overline{q}}=6$ equivalent $\overline{q}$-vectors in the BZ, over which we sum. 
$R_{\lambda \overline{q}_i}$ refers to a set of atomic displacements according to phonon mode $\lambda$ with momentum $\overline{q}_i$ (indicated with a vertical teal line in the phonon dispersion plot of Fig. \ref{fig:1}).
The quantity $\partial^2 \varepsilon^{(0)}_j(\omega)/\partial R_{\lambda \overline{q}_i}^2$ is evaluated by finite displacement from the equilibrium atomic positions.\footnote{The magnitude of the lattice displacement for the finite-difference derivatives was converged to just above the threshold of numerical noise. The harmonic behaviour of $\varepsilon (\omega)$ with respect to lattice displacements was numerically verified.
By varying the light polarisation directions, we confirm that the spectrum resulting from $\varepsilon^{\mathrm{st.,}(2)}_{\overline{q}2}(\omega)$ is $\pi/3$-periodic so that $\sum_i^{N_{\overline{q}}}[\dots ] \rightarrow 6[\dots]$.}
The last factor is the thermal average of the squared displacement of a quantum harmonic oscillator, given by $\sigma^2_{\lambda \overline{q}}=l^2_{\lambda\overline{q}}[2n_{B}(\Omega_{\lambda \overline{q}}, T)+1]$, with $n_B$ being the Bose-Einstein distribution for phonons and $l^2_{\lambda\overline{q}}=\hbar/(2M_{\lambda\overline{q}}\Omega_{\lambda\overline{q}})$.\footnote{$M_{\lambda \overline{q}}=\sum_{\kappa=\mathrm{N},\mathrm{B}} m_{\kappa} |e_{\kappa,\lambda \overline{q}}|^2$ is the generalised unit cell mass obtained from the normalisation condition of phonon eigenvectors $e_{\lambda \overline{q}}$ and atomic masses $m_B$ for boron and $m_N$ for nitrogen.}
Even though the function $\varepsilon^{\mathrm{st.,}(2)}_{\overline{q}}(\omega)$ includes contributions from the derivatives of both $T^S$ and $E^S$, for indirect transitions only one term survives in its imaginary part:
\begin{equation}\label{eq:ImTaylor}
\Im \frac{\partial^2 \varepsilon^{(0)}(\omega)}{\partial R_{\lambda \overline{q}}^2}\Bigr|_{\mathrm{eq}} = C \sum_{S^\prime} \frac{\partial^2 |T^{S^\prime}|^2 }{\partial R_{\lambda \overline{q}}^2}\Bigr|_{\mathrm{eq}}\Im\left\{\frac{1}{\hbar\omega-E^{S^\prime}+\mathrm{i}\eta}\right\},
\end{equation}
where $S^\prime$ restricts the sum to the $q=\overline{q}$ excitons.
This result allows us to reintroduce the phonon frequency dependence of $\varepsilon^{(2)}_{\overline{q}}(\omega) $ by imposing the correct energy conservation from perturbation theory and distinguishing between phonon emission ($\propto n_B+1$) and phonon absorption ($\propto n_B$): $[2n_B+1]/(\hbar\omega-E^{S^\prime}+\mathrm{i}\eta) \rightarrow 
[n_B+1]/(\hbar\omega-E^{S^\prime}-\hbar\Omega_{\lambda \overline{q}} + \mathrm{i}\eta) +  
n_B/(\hbar\omega-E^{S^\prime}+\hbar\Omega_{\lambda \overline{q}} + \mathrm{i}\eta).$
Finally, renaming the numerator between square brackets in Eq. \eqref{eq:taylor} (including the $1/2$ factors) as $|t^{\mathrm{static}}_{\lambda\overline{q}S^\prime}|^2$, since it represents the static formation probability of exciton $S^\prime$ mediated by a phonon mode $\lambda$ with momentum $\overline{q}$ and frequency $\Omega_{\lambda\overline{q}}$ , we obtain the final expression:
\begin{equation}\label{eq:indeps}
\varepsilon^{(2)}_{\overline{q}2}(\omega)=\sum_{\lambda S^\prime}|t^{\mathrm{static}}_{\lambda\overline{q}S^\prime}|^2 l^2_{\lambda\overline{q}}[n_B+1/2\mp 1/2] \delta(\hbar\omega-E^{S^\prime}\pm\hbar\Omega_{\lambda\overline{q}}).
\end{equation}
Here, the upper (lower) sign refers to the process of phonon absorption (emission).

Applying Eq.~(\ref{eq:indeps}) to the description of the process of exciton formation via photon absorption together with phonon emission at T = 0 K, computing the derivatives of $\varepsilon(\omega)$ with finite-difference DFT-BSE calculations in the non-diagonal supercell, we obtain the spectrum in Fig.~\ref{fig:2}.
It is possible to identify a multi-peak structure associated to the coupling of both the $i1$ and $i2$ excitons to all the in-plane phonon modes, with the higher-energy state $i2$ accounting for most of the oscillator strength.
The coupling of $i1$ and $i2$ with specific phonon modes depends on the light polarisation direction according to the symmetry selection rules detailed in Sec. VI of the SI.
Since the phonon frequencies are close to the energy difference between direct and indirect excitons, the phonon-assisted peaks are distributed around the brithtest direct exciton peak with most of the oscillator strength remaining in this narrow energy range. 
This result suggests that phonon-assisted absorption in hBN is at the origin of the fine structure observed around the brightest exciton peak in absorption experiments.\cite{Watanabe2004,Schue2018arXiv}
\begin{figure}[t!]
\includegraphics[width=0.9\columnwidth]{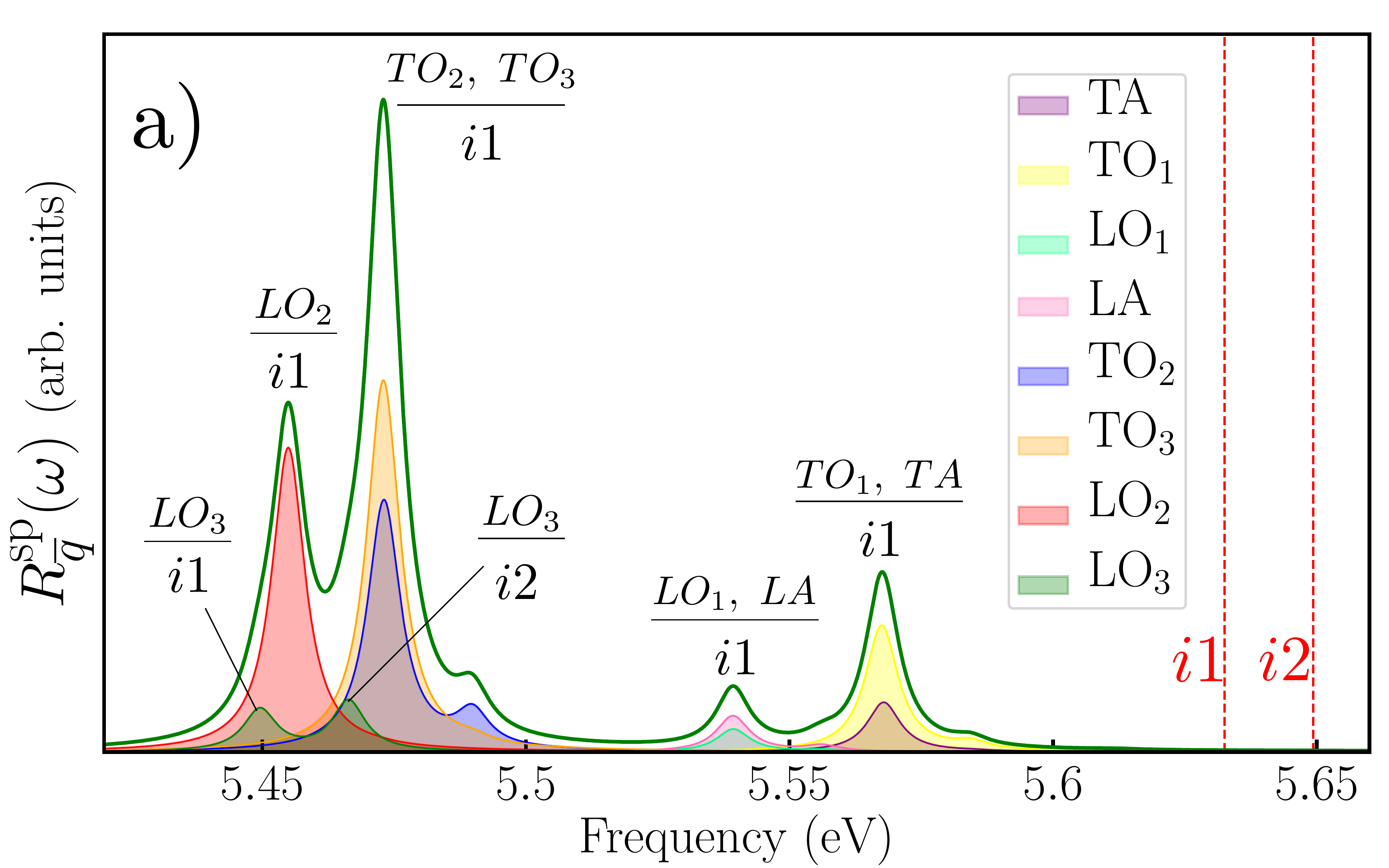}
\includegraphics[width=0.9\columnwidth]{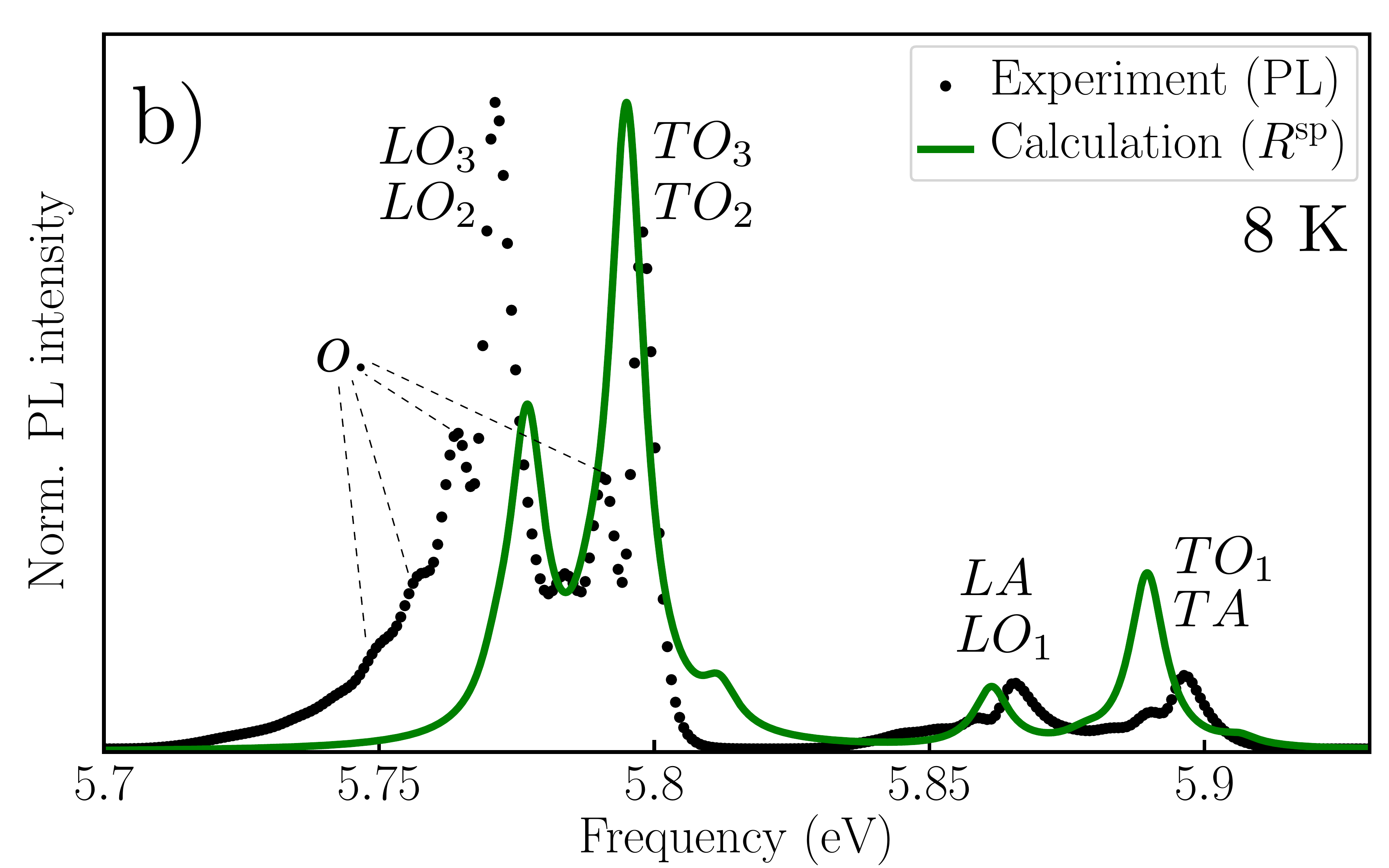}
\caption{Phonon-assisted emission in bulk hexagonal boron nitride. a) Spectral function $R^{\mathrm{sp}}_{\overline{q}}=\sum_{\lambda} R^{\mathrm{sp}}_{\lambda \overline{q}}$ (green solid line) at $8$ K. The $\lambda$-components of the spectrum, belonging to the different phonon modes, are also plotted in various colours. The exciton-phonon couplings are labeled. b) Comparison with the normalized experimental spectra (black dots) at $8$ K from Ref. [\onlinecite{exp_data}]. Notice that here our spectrum is blueshifted by $0.322$ eV to match the experimental one (the most prominent multi-phonon overtones are denoted as ``o.''). The temperature-dependent peak widths are described according to a linear model\cite{Rudin1990} (parameters taken from the experimental fit, see SI). The experimental excitonic temperature $T_{exc}=55$ K (see text) is used.\label{fig:3}}
\end{figure}

We now turn to the calculation of luminescence spectra. Here, we will concentrate on photon-emission assisted by phonon emission (whose intensity is $\propto n_B+1$ while the intensity of luminescence assisted by phonon absorption is $\propto n_B$, i.e., is negligible at small temperatures). The energy of an emitted photon differs from the one of an absorbed photon (indirect absorption with phonon emission) by twice the frequency of the phonon involved. We thus define $\varepsilon^{\mathrm{em}}_{\lambda \overline{q}}(\omega) = \varepsilon^{(2)}_{\lambda \overline{q}} (\omega - 2 \Omega_{\lambda \overline{q}})$.
This implies that the probabilities of exciton formation and annihilation are the same (i.e. detailed balance of phonon-assisted optical processes), a reasonable assumption in our scheme as we are computing the transition rates in the static approximation. 
Because of this, each phonon-assisted peak is mirrored with respect to the energy of the excitonic state involved.
Such approach, however, does not take into account the fact that the emission process should be proportional to the temperature-dependent occupations of the excitonic states, which leads to the reversal of the relative intensities of some peaks and to the quenching of others, when compared to $\Im \varepsilon^{\mathrm{em}}_{\lambda \overline{q}}(\omega)$.
In order to properly describe emission processes, we employ the Van Roosbroeck-Shockley (RS) relation to compute the spontaneous emission rate starting from the absorption coefficient,\cite{landsberg} by making two common experimental assumptions: (i) absorption and emission are in steady state, with the contribution from stimulated emission being very small; (ii) in steady state, the occupation functions of the excitonic states involved in recombination processes can be approximated with those at thermal equilibrium (Bose-Einstein distribution for excitons).
To apply the RS relation, originally derived in the context of direct independent-particle transitions,\cite{vRS} we have to extend it to excitons (see Sec.~VII of SI): In the low-density limit, this can be approximated by the Boltzmann factor $N_B(\Delta E_{i1,i}) = \mathrm{e}^{-(E^i - E^{i1})/k_BT_{\mathrm{exc}}}$ ($E^{i1}$ being the energy minimum of the exciton dispersion curve and $E_i$ the energy of any exciton $i$).
\begin{equation}\label{eq:RS}
R^{\mathrm{sp}}_{\overline{q}}(\omega)= \sum_\lambda \frac{\omega(\omega + 2\Omega_{\lambda \overline{q}})^2}{\pi^2 \hbar c^3 } n_r(\omega) \Im \{ \varepsilon^{\mathrm{em}}_{\lambda \overline{q}}(\omega)\} N_B.
\end{equation} 
It is sufficient to compute the refractive index from $\varepsilon^0(\omega)$, which completely determines its slow decay at low frequencies. 
Because of the large energy difference between $E^{i1}$ and the main direct peak, the latter will always be suppressed by the Boltzmann factor $N_B$ up to room temperature, and therefore it should not be seen in a PL experiment.
The results for $R^{\mathrm{sp}}_{\overline{q}}(\omega)$ are plotted in Fig. \ref{fig:3}a.
The energy differences between excitonic levels ($\Delta E_{i1,i2} =17$ meV) and the phonon frequencies (from $50$ to $200$ meV) are large, giving rise to a well-spaced peak structure that can be easily resolved.
We notice two separated groups of features which are clearly seen in experiment (Fig. \ref{fig:3}b, black dots): one at higher energy generated by low-frequency phonon emission, and the other at lower-energy due to high-frequency optical phonons.
The emission spectrum is almost completely dominated by the lowest-bound exciton $i1$ since the occupation factor quenches most of the peaks related to $i2$.
However, experiments have shown\cite{CassaboisG.2016} that in bulk hBN the excitonic temperature $T_{exc}$ that goes into the Boltzmann factor $N_B$ is greater than the lattice temperature $T_L$.
If we set $T_{exc}=T_L=8$ K in $R^{\mathrm{sp}}_{\overline{q}}$, we obtain only the peaks coming from $i1$, while setting $T_{exc}$ to the experimental value of $55$ K (as in Fig. \ref{fig:3}a) leads to the appearance of the quenched peaks from $i2$.
On the experimental side, the peaks due to high-frequency modes are clearly visible in Fig. \ref{fig:3}b at $8$ K: the separation between these peaks thus corresponds to the separation between the LO and TO modes at point $\overline{q}$ in the theoretical calculations, while in experiment it corresponds to the splitting at point $\overline{q}+\Delta q$, where $\Delta q$ is the error committed by approximating the `true' $q$-point $|T_1 -M|$ with $\overline{q}=|K-M|$.
We attribute to this discrepancy the difference of $\sim 80$ meV between the theoretical and experimental peak separations.\footnote{Our error in $q$-space, possibly affecting the positions of the peaks related to phonon modes with a steep dispersion curve, is approximately $12$ \% of $\Gamma$K, with $\Delta q=0.01$ \AA$^{-1}$.}
As expected, we cannot capture the various satellite peaks in Fig. \ref{fig:3}b, which experimentally are assigned to multi-phonon processes involving zone-center shear phonon modes.\cite{Vuong-overtones}
In our results, the quasi-degenerate transverse modes TO$_2$ and TO$_3$ couple with similar strength to $i1$, whereas the longitudinal mode LO$_3$ has a higher frequency than LO$_2$ and a much weaker coupling to $i1$ (in the absorption case, the coupling of mode LO$_3$ with $i2$ is instead the strongest one).
This leads to a discrepancy in the relative intensities of peaks LO$_2$/LO$_3$ and TO$_2$/TO$_3$ between theory and experiment.
We consider it possible that the non-negligible sum of the overtone peaks to the main LO$_2$/LO$_3$ peak is responsible for such inversion in relative intensity.

In conclusion, we have presented a theoretical and computational approach to phonon-assisted absorption and emission in bulk hBN where very strong excitonic effects are present. Using a supercell for finite-q excitons and a finite-difference method for exciton-phonon coupling,
we are able to reproduce the single-phonon-assisted spectral features obtained in photoluminescence experiments. The structure of the emission spectrum can be understood in terms of a Davydov pair of finite-q excitons coupling with different strengths to the in-plane phonon modes.
The method can be applied to other layered materials with indirect gap, e.g. bilayers of transition metal dichalcogenides or to the case of monolayer WSe$_2$ whose optical gap, previously assumed to be ``direct'', might in fact be ``indirect''.\cite{Wei-Ting2017}
Moreover, it will stimulate further investigations directed toward a full many-body perturbation theory of exciton-phonon coupling, as well as to sound and reliable approximations applicable to  calculations of more complex materials.

F.\,P. and L.\,W. acknowledge support from the FNR, Luxembourg (Projects EXCPHON/11280304 and INTER/ANR/13/20/NANOTMD, respectively). A.\,M.-S. acknowledges the Juan de la Cierva (Grant IJCI-2015-25799) program (MINECO, Spain). H.\,M. acknowledges the F.R.S.-FNRS through the PDR Grants HTBaSE (T.1071.15). 
We acknowledge F.\,Giustino for suggesting the use of the RS relation and M.\,Zacharias, T.\,Galvani, M.\,Barborini, and S.\,Reichardt for stimulating discussions. 
We are also indebted to T.\,Q.\,P.\,Vuong, G.\,Cassabois and B.\,Gil for critical observations about the preliminary results.

\bibliography{excphon}

\end{document}


\title{Supplementary material -- Exciton-phonon coupling in the UV absorption and emission spectra of bulk hexagonal boron nitride}

\author{Fulvio Paleari}
\affiliation{Physics and Materials Science Research Unit, University of Luxembourg, 162a avenue de la Fa\"iencerie, L-1511 Luxembourg, Luxembourg}

\author{Henrique P. C. Miranda}
\affiliation{Physics and Materials Science Research Unit, University of Luxembourg, 162a avenue de la Fa\"iencerie, L-1511 Luxembourg, Luxembourg}
\affiliation{Institute of Condensed Matter and Nanosciences, Universit\'{e} catholique de Louvain, Chemin des \'etoiles 8, bte L7.03.01, 1348, Louvain-la-Neuve, Belgium}

\author{Alejandro Molina-S\'{a}nchez}
\affiliation{Institute of Materials Science (ICMUV), University of Valencia, Catedr\'{a}tico Beltr\'{a}n 2, E-46980 Valencia, Spain}

\author{Ludger Wirtz}
\affiliation{Physics and Materials Science Research Unit, University of Luxembourg, 162a avenue de la Fa\"iencerie, L-1511 Luxembourg, Luxembourg}

\date{\today}

\maketitle

\section{Methods}
\begin{figure}[b]
\includegraphics[width=\columnwidth]{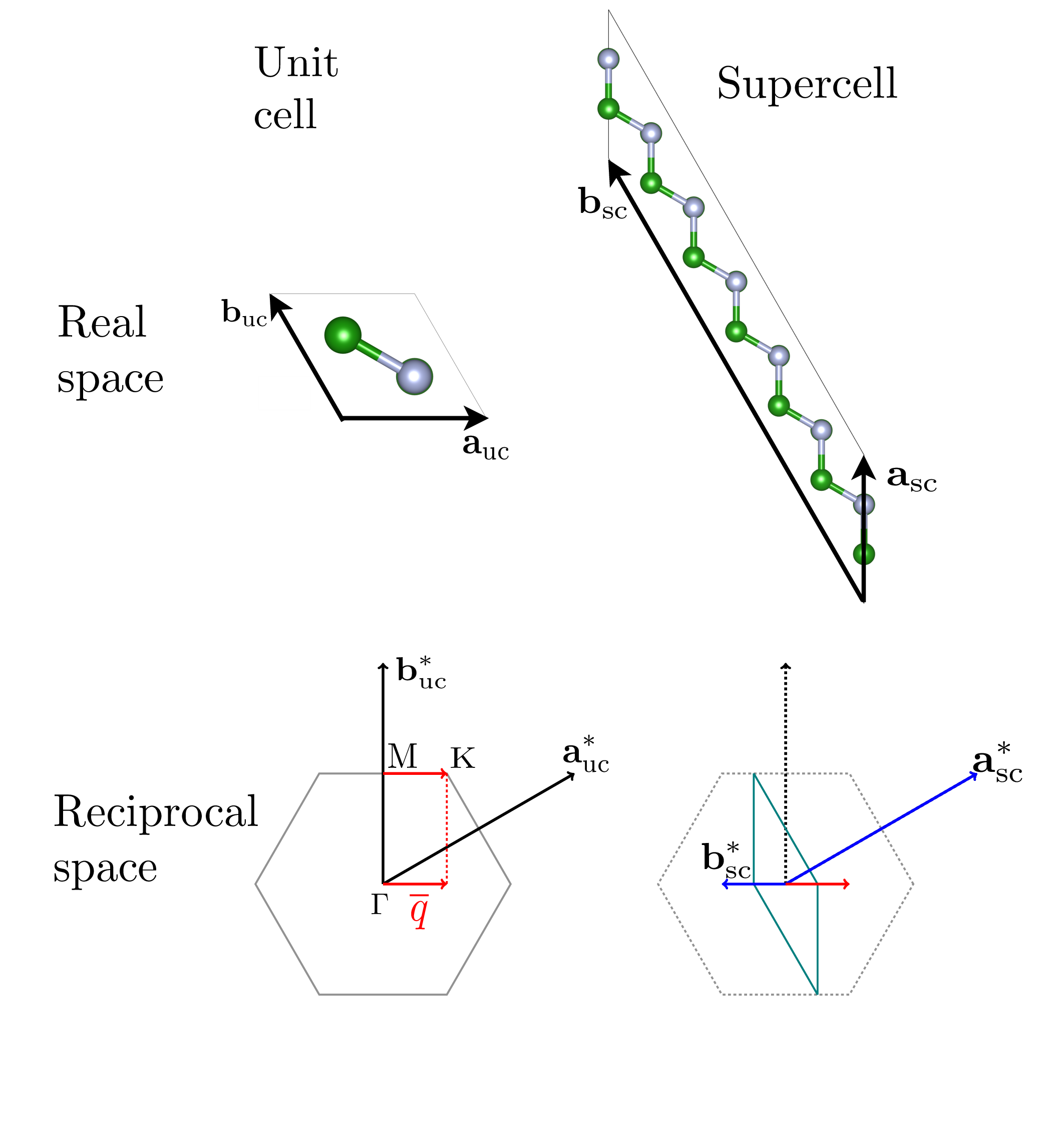}
\caption{Top: representations of the hBN unit cell (uc) and non-diagonal supercell (sc) used in our calculations (boron is green, nitrogen is gray). Bottom: schemes of the reciprocal-space BZ in the two cases (the BZ borders are in gray and teal, respectively), showing that in the first Brillouin zone of the supercell, the momentum $\overline{q}$ is folded back onto the $\Gamma$ point.}
\label{sfig:1a}
\end{figure}

The non-diagonal supercell\cite{lloyd} used in our calculations is represented in Fig. \ref{sfig:1a} and compared with the unit cell.
The comparison between the respective reciprocal-space Brillouin zones (BZ) is also shown, emphasising the folding of the q-point $\overline{q}$ (given in fractional coordinates as $(1/3,-1/6,0)$) into $\Gamma$ in the supercell case (note that both $\overline{q}$ and $-\overline{q}$ are folded onto $\Gamma$ in this way).

\begin{figure}[t]
\includegraphics[width=\columnwidth]{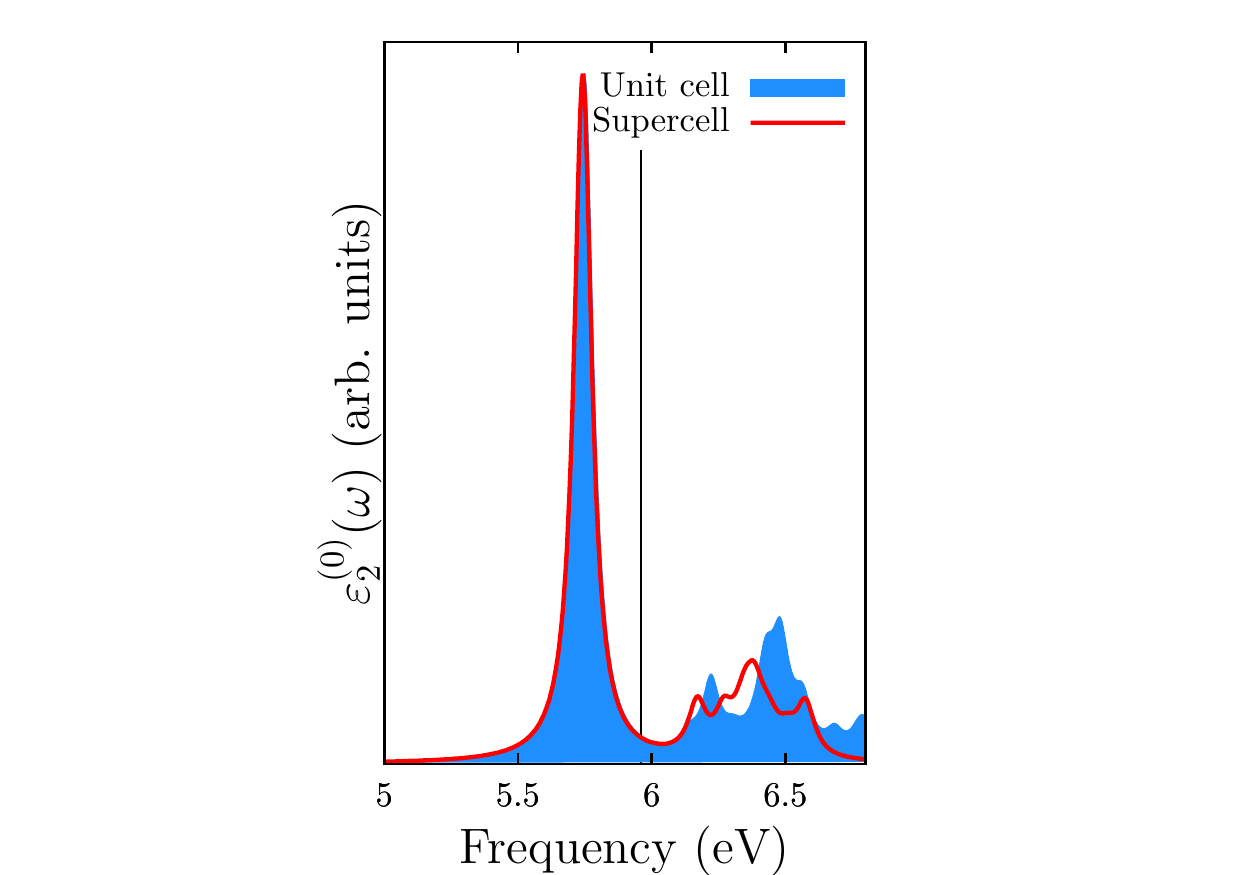}
\caption{The imaginary part of the dielectric functions are plotted for the hBN unit cell (blue, DFT-GW-BSE calculation on fine $k$-point grid) and non-diagonal supercell (red, DFT-scissor-BSE calculation on coarse $k$-point grid). Only the first $480$ states are included in the iterative solution of the BSE in the supercell. The black vertical line is at the energy of the indirect quasiparticle band gap. Both spectra have a broadening parameter of $0.04$ eV. Difference in the spectra (due to finite k-point samplings) only occur in the energy region above the band gap. In the relevant energy region below the band gap, both spectra are identical.}
\label{sfig:1}
\end{figure}

The DFT ground state calculations were performed with Quantum ESPRESSO\cite{QE} using LDA\cite{lda} norm-conserving pseudopotentials\cite{FUCHS199967} with a kinetic energy cutoff of $110$ Ry. 
The phonon frequencies and eigenmodes were computed with density functional perturbation theory (DFPT),\cite{dfpt} using a $q$-point grid sampling of $18 \times 18 \times 6$.
The G$_0$W$_0$ and semi-self-consistent GW$_0$ (sscGW$_0$) corrections to the band energies were obtained with the YAMBO code\cite{Marini2009} for the unit cell, using the plasmon-pole approximation for the dynamical screening.
The direct and indirect gaps were converged with a $18\times 18 \times 6$ $k$-point grid, summing $160$ and $280$ states for the screening function and the Green's function, respectively.
The corrections were computed for the last $4$ valence bands and the first $6$ conduction bands.
The sscGW$_0$ amounts to an additional opening of the band gap by $0.22$ eV with respect to the single G$_0$W$_0$ calculation.\cite{multiBN}
The fully converged result was subsequently used to construct a $k$-dependent scissor operator in such a way that, when applied to the supercell, it would yield exactly the same optical absorption spectrum as the unit cell (here we neglect the changes in the GW corrections due to lattice displacements).
The exciton energies $E^s$ and wave functions $\Phi^S$ were computed with YAMBO by solving the Bethe-Salpeter equation (BSE) with RPA static screening and the Tamm-Dancoff approximation, including the GW-corrected band structures.
The Bethe-Salpeter equation in the supercell is solved iteratively in the YAMBO code using the SLEPC library\cite{Hernandez2005} for the first $600$ eigenvalues and eigenvectors.
The macroscopic dielectric function $\varepsilon^{(0)}(\omega)$ was computed using the modified response function where the long-range Fourier component is removed.\cite{Onida2002}
Its resonant part can be written as:
\begin{equation}\label{eq:exceps}
\varepsilon^{(0)}(\omega)= 1 - \frac{8\pi}{N_k V_R}\sum_S \frac{\lvert T^S\rvert^2 }{\hbar\omega-E^S+\mathrm{i}\eta}.
\end{equation}
Here, $N_k$ is the number of k-points in the Brillouin zone sampling and $V_R$ the BZ volume.
The term in the numerator, $T^S=\sum_{kcv} \Phi^S_{cvk}\bra{ck}\mathbf{e}\cdot\hat{\mathbf{D}}\ket{vk}$, is the oscillator strength of exciton $S$.
It is a linear combination of optical matrix elements in the dipole approximation ($\mathbf{e}$ is the light polarisation direction), describing direct transitions from an occupied state $v$ to an unoccupied one $c$, weighted by the corresponding components of the excitonic wave function. 
In the unit cell, a reasonably converged calculation of the static screening can be obtained by considering a $18\times 18 \times 6$ $k$-point grid and summing $250$ bands.
However, the energy window close to the absorption edge is already converged with a $12\times 12 \times 4$ sampling.
As the non-diagonal supercell contains $6$ times the atoms of the unit cell, the convergence parameters were changed accordingly, using a $12 \times 2 \times 4$ $k$-point grid, and including enough states in the Bethe-Salpeter kernel to span the transition energy region relevant for the absorption edge. 
The static screening was computed summing $1.2\cdot 6 \cdot 250$ bands. The factor $1.2$ is a safety margin to account both for the folded bands from the zone edge and for spurious finite-$q$ bands (see Fig.~\ref{sfig:1} for a comparison of the optical absorption spectra $\Im \varepsilon^{(0)}(\omega)$ between unit cell and supercell).
The normalised phonon displacements (rescaled by the square root of the atomic masses) were used to compute the finite-difference derivatives.
A global scaling factor was multiplied by the displacements and converged to $0.0025$ \AA, which is just above the threshold of numerical noise. 
The harmonic behaviour of $\varepsilon (\omega)$ with respect to the scaling factor was numerically verified.

\begin{figure}[ht!]
\includegraphics[width=\columnwidth]{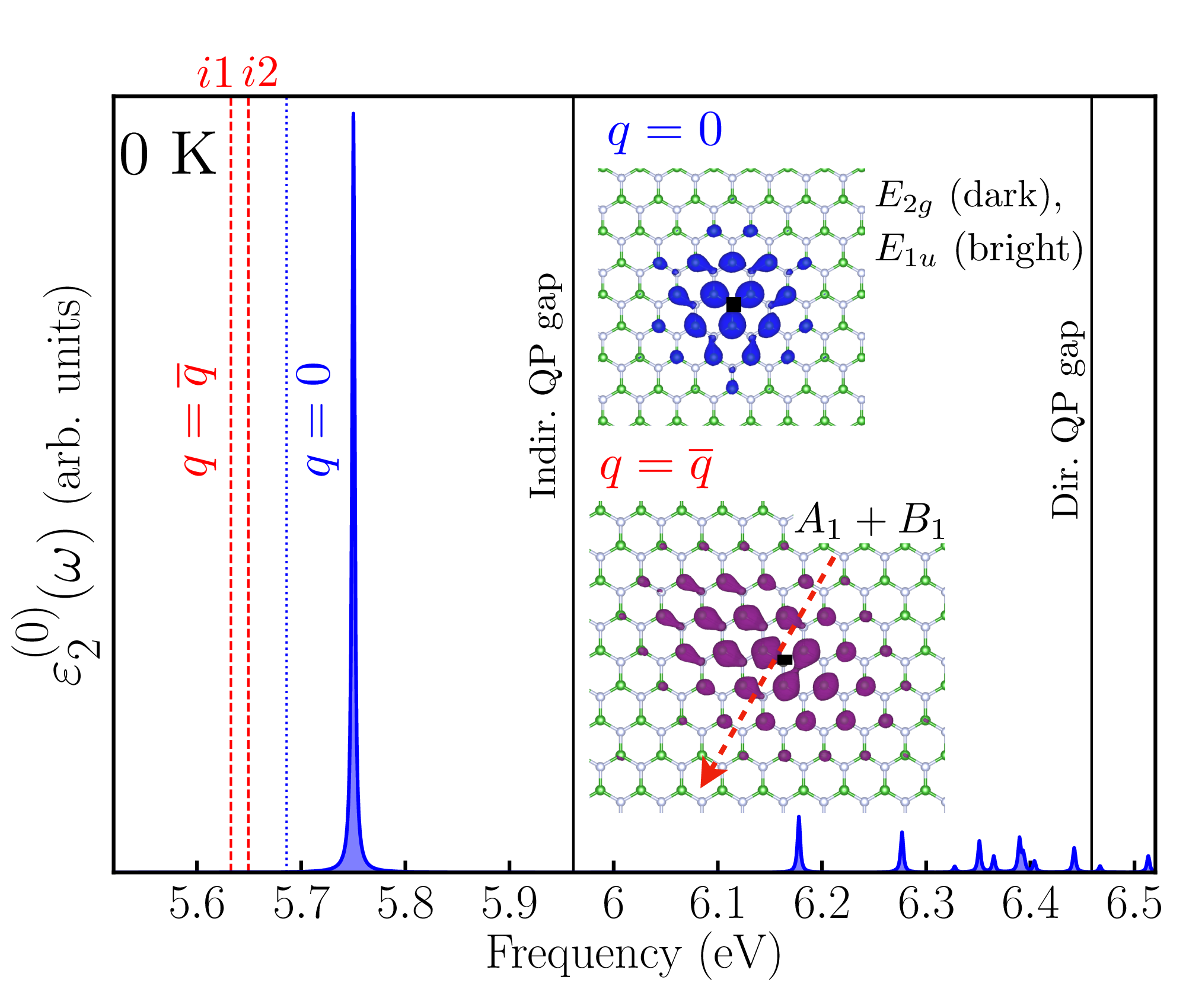}
\caption{Equilibrium supercell calculation. The imaginary part of the dielectric function including only transitions at $q=0$ ($\Im \varepsilon^{(0)} (\omega)$ in the text) is shown in blue. The peak broadenings are set to $1.5$ meV for the temperature of $0$ K. The vertical black lines denote the quasiparticle indirect and direct gaps, while the dotted blue line indicates the position of the dark E$_{2g}$ exciton with $q=0$. The dashed red lines indicate the positions of the A$_1$ and B$_1$ excitons at $\overline{q}$ (labeled $i1$ and $i2$, respectively). The excitonic wave functions are plotted (intensities only) in the insets: blue for the $q=0$ pair, and purple for $q=\overline{q}$ ones (the dashed red arrow labeling the $\overline{q}$ direction). Here the hole (black square) is fixed on top of a nitrogen atom, and the resulting electronic density is displayed.\label{sfig:2a}}
\end{figure}
\section{Equilibrium results on the supercell}
The result of a BSE calculation in the supercell with atoms fixed at their equilibrium positions is shown in Fig. \ref{sfig:2a}.
The energy separations between direct and indirect band gaps, as well as with low-lying $q=0$ and finite-$q$ excitonic states, are emphasized by vertical lines. 
The first $q=0$ exciton at $5.7$ eV (dashed blue vertical line) is dark, while the second one at $5.75$ eV (main blue peak) is optically active. 
The intensities of their wave functions $\Psi (\mathbf{r_e},\mathbf{r_h})$ are almost identical. 
The one of the bright exciton is represented in blue in the inset of Fig. \ref{sfig:2a}. Both electron and hole mostly lie in the same layer.
Two finite-$q$ excitonic states appear below the direct ones: they are labeled $i1$ and $i2$ (dashed red lines).
The wave function intensities of the two indirect excitons are very similar.
One of them is displayed in purple in the second inset of Fig. \ref{sfig:2a}, showing again a mostly planar distribution.
Although the wave function in the fixed-hole representation looks approximately distorted along the armchair lattice direction, the full wave function is actually symmetric with respect to the zigzag direction (parallel to the $\overline{q}$-vector) upon rotation along the principal axis of the $C_{2v}$ symmetry group which is oriented in-plane along the $\overline{q}$ direction (see also Section \ref{s:group}). 

\section{Position of the direct band gap}
The direct quasiparticle band gap in bulk hBN is traditionally identified at the so-called $T_1$ point.\cite{Arnaud2006}
From Fig. 1 of the main text, we see that this point lies close to K, along the $\Gamma$K symmetry line (to be precise: $|\mathbf{T_1}-\mathbf{K}|\simeq 1/6|\mathbf{K}-\boldsymbol\Gamma|$).
However, there are other three points that give a comparable band gap: M, $T_2$ (along the MK line: $|\mathbf{T_2}-\mathbf{K}|\simeq1/3|\mathbf{K}-\mathbf{M}|=1/6|\mathbf{K}-\boldsymbol\Gamma|$), and H, the high-symmetry point directly above K along in the out-of-plane direction.
In our GW calculations, the band gaps at these points lie in a small $0.1$ eV energy interval (with $E_g^H<E_g^{T_2}<E_g^{M}<E_g^{T_1}$), which is the accuracy of the GW method.
Since we know that GW underestimates the true quasiparticle corrections in hBN, we have to assume that the relative energy differences between these band gaps may change with more refined approximations and/or more accurate calculations.
In the main text, we ``average'' the true position of the top of the valence band from ``around" K to exactly K, by taking $\overline{q}=|\mathbf{K}-\mathbf{M}| = 0.5|\mathbf{K}-\boldsymbol\Gamma|$ as the momentum transfer corresponding to the indirect gap.

\section{Model for line broadenings}
In our calculations, the temperature-dependent exciton lifetime, which is related to the imaginary part of the exciton-phonon self-energy and inversely proportional to the line broadenings $\eta$, remains an empirical parameter. 
As the imaginary part of our response functions intrinsically gives a Lorentzian shape for single peaks ($\mathrm{Im} [ \omega - E - \mathrm{i} \eta ] ^{-1}$), we focus on the range of temperatures in which the experimental phonon-assisted peaks can also be reasonably described with a Lorentzian broadening (from $0$ to $100$ K).\cite{exp_data}
We use a linear model where the line broadening $\eta$ is given by\cite{Rudin1990}
\begin{equation*}
\eta = \Gamma_0 + aT +b N_{BE}(T),
\end{equation*}
where $N_{BE} = [\mathrm{e}^{E_O/kT}-1]^{-1}$ is the Bose-Einstein distribution.
The values of the parameters are taken from the experimental fit in Ref. [\onlinecite{exp_data}]: $\Gamma_0 = 3$ meV, $a = 0.1$ meV/K, $b = 150$ meV, and $E_O = 25$ meV.

\begin{figure*}[ht]
\includegraphics[width=0.8\textwidth,trim={0cm, 0cm, 0cm, 0cm}]{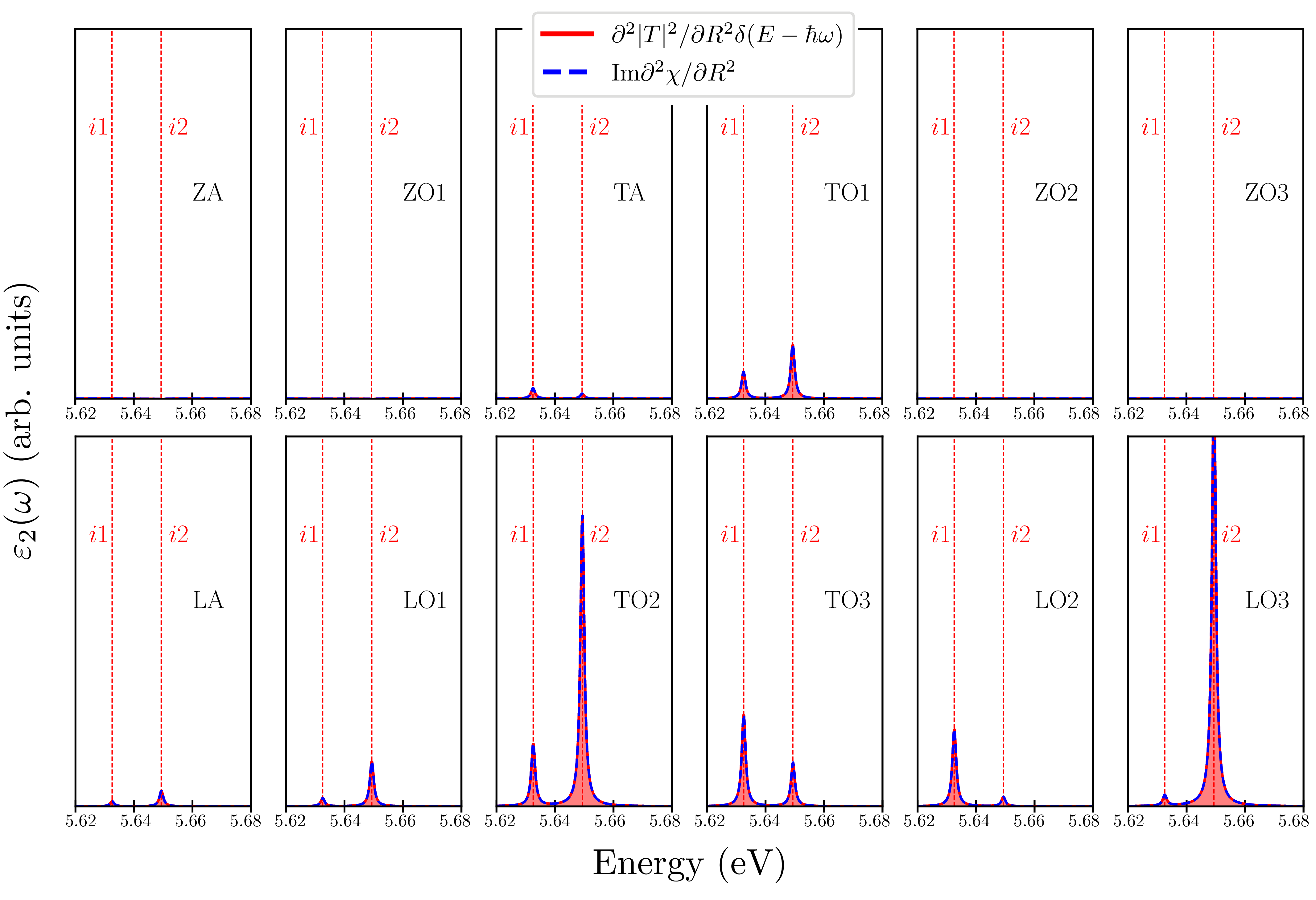}
\caption{Static exciton-phonon couplings between the $i1$ and $i2$ excitons and the 12 different phonons at $\overline{q}=\frac{1}{2}|\Gamma K|$. We compare results obtained from the full second derivative of the response function (dashed blue lines) and from the second derivative of the oscillator strengths (full red lines), see Eq. \eqref{eq:d2}. An average over polarization of the incoming light and sum over the 6 equivalent directions of phonons with wave vector $|\overline{q}|$ is performed.}
\label{sfig:3}
\end{figure*}

\section{Form of the finite-difference response function}
Here we give the full derivation of Eqs. (2) and (3) of the main text.
Let us consider the part of the response function due to excitonic state $S$ (see Eq. \eqref{eq:exceps} for the full response function):
\begin{equation}\label{eq:chi}
\chi^S_R (\omega ) = \frac{\lvert T^S_R \rvert^2}{E^S_R-\hbar\omega+\mathrm{i}\eta},
\end{equation}
where the subscript $R$ denotes a parametric dependence on lattice displacements, i.e. $\chi^S_0$ is the frozen-atom response function (we take for simplicity $\eta$ to be independent of $R$, an assumption that does not affect the validity of the results).
We want to make the Taylor expansion of Eq. \eqref{eq:chi} up to second order in the lattice displacements, therefore as an initial step we need to compute its first derivative and evaluate it at the equilibrium atomic positions:
\begin{widetext}
\begin{equation}
\frac{\partial \chi^S_R (\omega )}{\partial R}\Bigr|_{R=0} = \frac{\partial |T^S_R|}{\partial R}\Bigr|_{R=0} 2|T^S_{R=0}| [E^S_{R=0}-\hbar\omega+\mathrm{i}\eta]^{-1}+\frac{\partial [E^S_R-\hbar\omega+\mathrm{i}\eta]^{-1}}{\partial R}\Bigr|_{R=0}\lvert T^S_{R=0} \rvert^2.
\end{equation}
\end{widetext}
At this point we notice that the oscillator strengths of any finite-$q$ excitons for optical absorption must be zero because of momentum conservation. In the case of a supercell, it means that the excitons being folded into $\Gamma$ from a different point in the unit cell are dark if the atoms are clamped at the equilibrium positions.
If we label excitons belonging to such subset with $S^\prime$, this means that $T^{S^\prime}_{R=0}=0$ and therefore $\partial \chi^{S^\prime}_R (\omega )/\partial R\Bigr|_{R=0}=0$.
This is explicitly confirmed numerically by our finite-difference calculations.
This argument is analogous to the one often used in the case of optical absorption in the independent-particle model for the vanishing of the dipole optical matrix elements below the direct band gap.\cite{hbb,Marios2}

The same argument applied to the second derivative of $\chi^S_R$ leads to the vanishing of the terms containing derivatives of the exciton binding energy.
The only term that remains is the one containing the second derivative of $T^{S^\prime}_R$:
\begin{equation}\label{eq:d2}
\frac{\partial^2 \chi^{S^\prime}_R(\omega)}{\partial R^2}\Bigr|_{R=0} = \frac{\partial^2 \lvert T^{S^\prime}_R \rvert^2}{\partial R^2}\Bigr|_{R=0}[E^{S^\prime}_{R=0}-\hbar\omega+\mathrm{i}\eta]^{-1},
\end{equation}
which corresponds to Eq. (3) of the main text.
This derivative is evaluated numerically using the finite-difference expression
\begin{displaymath}
\partial^2 \chi_R /\partial R^2 (\omega) \approx [\chi(\Delta \mathbf{R} ; \omega) -2\chi_0(\omega) + \chi(-\Delta \mathbf{R} ; \omega)
]/\Delta \mathbf{R}^2.
\end{displaymath}
 The results, displayed in Fig. \ref{sfig:3} for each phonon mode, confirm the equivalence of the two sides of Eq. \eqref{eq:d2}.
 This means that, numerically, we can obtain the exciton-phonon coupling for the calculation of phonon-assisted absorption/emission both through a finite-difference calculation of the whole response function or through a finite-difference calculation of just the excitonic oscillator strength.

\section{Exciton symmetry and coupling with phonons}\label{s:group}
\begin{figure*}[ht]
\includegraphics[width=0.8\textwidth]{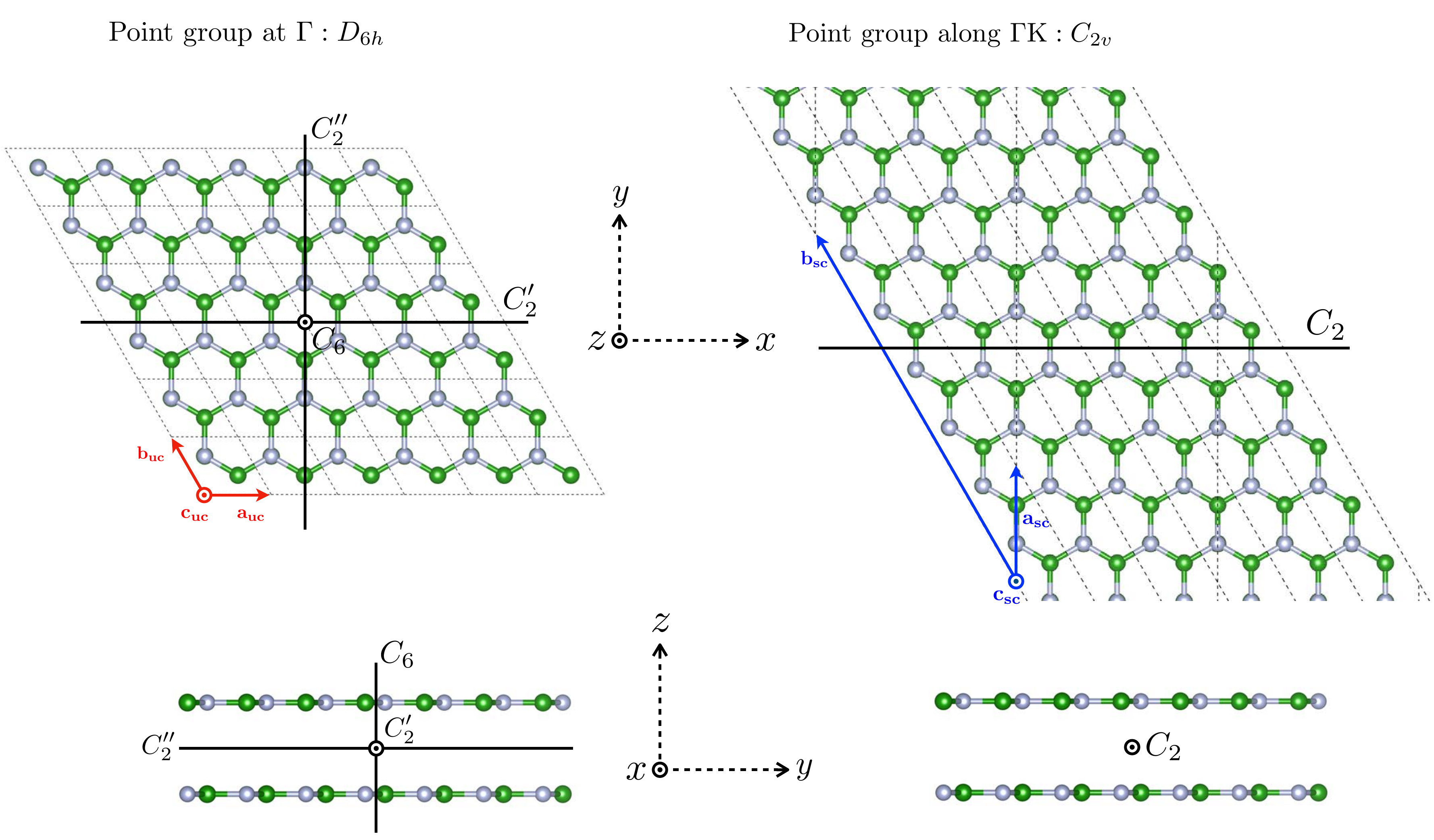}
\caption{Symmetries at $\Gamma$ and $\overline{q}$. Top and side views of the bulk hBN lattice. Boron (nitrogen) atoms are in green (gray). The borders of the repeated unit cells are shown with dashed black lines. Left: original unit cell for the hexagonal lattice  ($D_{6h}$ point group) and lattice vectors. Right: non-diagonal supercell used in the calculations, folding point $\overline{q}$ ($C_{2v}$ point group) at $\Gamma$, and lattice vectors. The axes of rotation corresponding to the symmetries of the $q$ points of the systems are shown with solid black lines. }
\label{sfig:4}
\end{figure*}

\begin{table}
\begin{center}
\begin{ruledtabular}
\begin{tabular}{c|cccc}   
$D_{6h}$ \ & \ $E$ \ & \ $C^{\prime}_2(x)$ \ & \ $\sigma_h(xy)$  \ & \  $\sigma_d(xz)$ \\
\hline
$A_{1g}$ & $+1$ & $+1$ & $+1$ & $+1$ \\
$A_{2u}$ & $+1$ & $-1$ & $-1$ & $+1$ \\
$E_{1g}$ & $+2$ & $0$ & $-2$ & $0$  \\
$E_{2g}$ & $+2$ & $0$ & $+2$ & $0$ \\
$E_{1u}$ & $+2$ & $0$ & $+2$ & $0$ \\
$E_{2u}$ & $+2$ & $0$ & $-2$ & $0$\\
\end{tabular}
\end{ruledtabular}
\caption{Partial character table for point group $D_{6h}$.\label{t:D6h}}
\end{center}
\end{table}
\begin{table}
\begin{center}
\begin{ruledtabular}
\begin{tabular}{c|cccc}   
$C_{2v}$ \ & \ $E$ \ & \ $C_2(x)$ \ & \ $\sigma_v(xy)$  \ & \  $\sigma_v(xz)$ \\
\hline
$A_1$ & $+1$ & $+1$ & $+1$ & $+1$ \\
$A_2$ & $+1$ & $+1$ & $-1$ & $-1$ \\
$B_1$ & $+1$ & $-1$ & $+1$ & $-1$ \\
$B_2$ & $+1$ & $-1$ & $-1$ & $+1$\\
\end{tabular}
\end{ruledtabular}
\caption{Character table for point group $C_{2v}$.\label{t:C2v}}
\end{center}
\end{table}
\begin{table}
\begin{center}

\begin{tabular}{ccc}   
$D_{6h}$ \ & \ & \ $C_{2v}$ \\
\hline
$C^{\prime}_2$ & $\rightarrow$ & $C_2$ \\
$\sigma_h(xy)$ & $\rightarrow$ & $\sigma_v(xy)$ \\
$\sigma_d(xz)$ & $\rightarrow$ & $\sigma_v(xz)$ \\
\end{tabular}

\caption{Connection between the elements of $C_{2v}$ and $D_{6h}$.\label{t:conn}}
\end{center}
\end{table}

\begin{table}
\begin{center}

\begin{tabular}{c|c|c}   
Mode \ & \ Symm. \ & \ Freq. (meV) \\
\hline
\hline
LO$_3$ & $A_1$ & $183.00$ \\
LO$_2$ & $B_1$ & $177.63$ \\ 
\hline
TO$_3$ & $A_1$ & $159.58$ \\
TO$_2$ & $B_1$ & $159.48$ \\
\hline
LA & $A_1$ & $93.33$ \\
LO$_1$ & $B_1$ & $93.22$ \\
\hline
ZO$_3$ & $A_2$ & $92.47$ \\
ZO$_2$ & $B_2$ & $87.53$ \\
\hline
TO$_1$ & $A_1$ & $65.10$ \\
TA & $B_1$ & $64.72$ \\
\hline
ZO$_1$ & $A_2$ & $22.25$ \\
ZA & $B_2$ & $21.54$ \\
\end{tabular}

\caption{Symmetry of the phonon modes at $\overline{q}$. The modes are listed as Davydov pairs in order of increasing frequency with the lowest-frequency mode at the bottom (compare Fig. (1) in the main text).\label{t:phonons}}
\end{center}
\end{table}
\begin{figure}[ht]
\includegraphics[width=0.8\columnwidth]{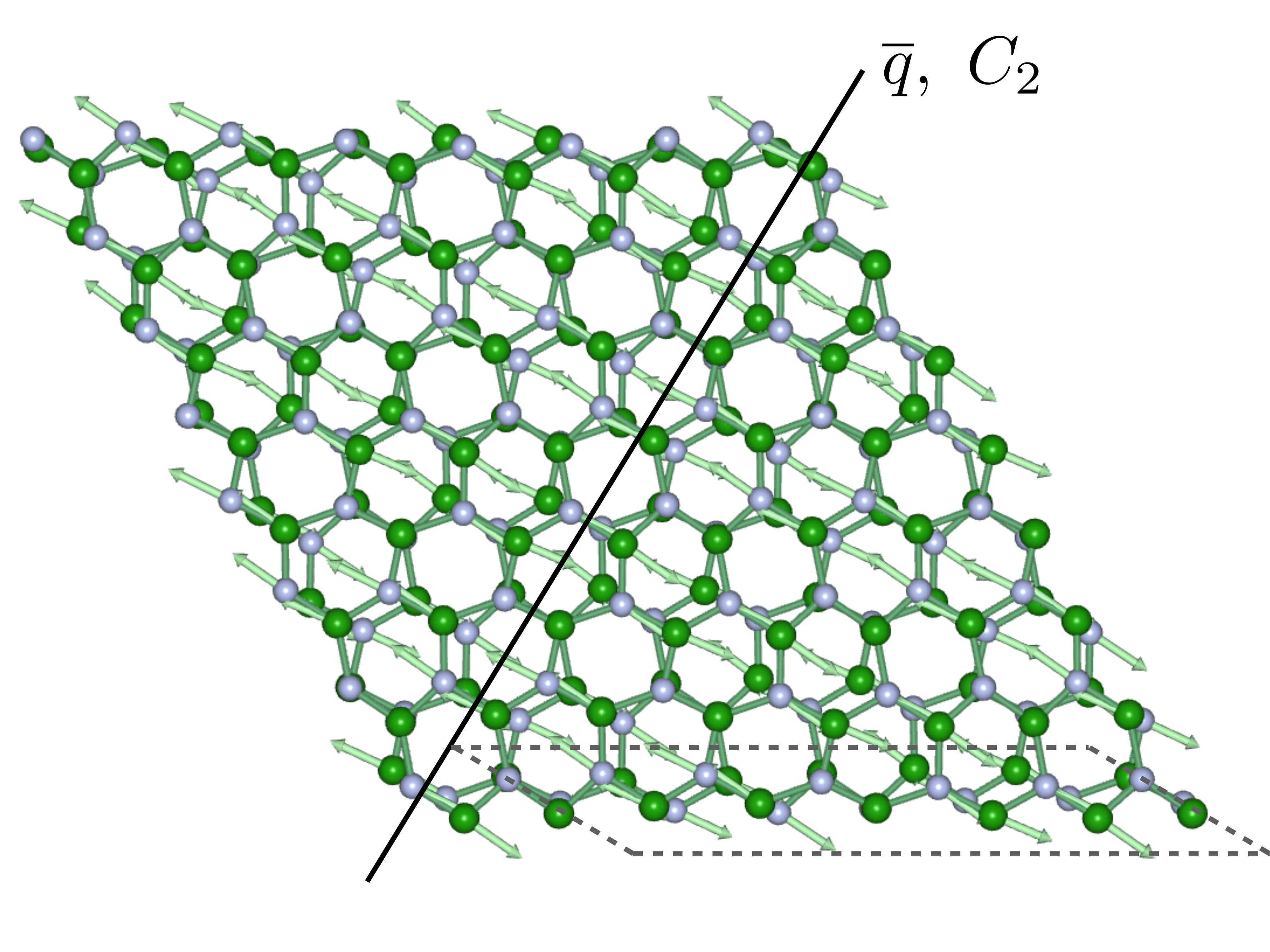}
\caption{Top view of the atomic displacements and forces for the TO$_1$ phonon mode at one of the six equivalent $\overline{q}$-vectors/C$_2$ rotation axes (solid black line). The corresponding non-diagonal supercell is represented in dashed gray lines.}
\label{sfig:4b}
\end{figure}

\subsection{Symmetry analysis with group theory}
In bulk hBN, the point group (including the non-symmorphic point symmetry operations of the space group) is $D_{6h}$. It contains $24$ symmetry operations grouped in $12$ classes. It is the group that is also used for the characterizations of perturbations (such as phonons and excitons) of hBN with zero wave-vector (corresponding to the high-symmetry point $\Gamma$).
In Table \ref{t:D6h} we report a subsection of the character table focusing on the operations and representations of interest to us. 
In Fig. \ref{sfig:4} (left panel) our choice for the Cartesian axes and for the lattice vectors is reported and the three rotation axes belonging to $D_{6h}$ are drawn. 
Recall that bulk hBN displays AA$^\prime$ stacking (two inequivalent layers per unit cell): therefore many of the symmetry operations are non-symmorphic.
The point group for the symmetry analysis of a perturbation with finite wave vector $\overline{q}$ is a subset of the one at $\Gamma$.
In the case of $\overline{q}$ lying on the $\Gamma$K line, the point symmetry group is $C_{2v}$ with $C_{2v}\subset D_{6h}$.
The character table of $C_{2v}$ is provided in Table \ref{t:C2v}.
In the right panel of Fig. \ref{sfig:4}, showing the crystal lattice as repetitions of the non-diagonal supercell used in our calculations, the only rotation axis of $C_{2v}$ is drawn.
This axis runs along the zigzag direction, and it is identified by checking the rotational symmetry of the phonon modes, as shown in Fig. \ref{sfig:4b}.\footnote{\label{note1}We used the following online tool to visualise the phonon dispersion and lattice displacements corresponding to different $q$-vectors: \url{http://henriquemiranda.github.io/phononwebsite/phonon.html}}
From this we see that the $C_2$ rotation in $C_{2v}$ coincides with the $C^{\prime}_2$ rotation in $D_{6h}$, and we use this to make a connection between the elements of $C_{2v}$ and $D_{6h}$, shown in Table \ref{t:conn}.
The dipole operator transforms as the $[x,y,z]$ vector and belongs to representations $E_{1u}[x,y]+A_{2u}[z]$ for $D_{6h}$ and $A_1[x]+B_1[y]+B_2[z]$ for $C_{2v}$. The in-plane component of the dipole transforms accordingly as $E_{1u}$ and $A_1+B_1$, respectively.

Let us first analyse the two excitons (one dark, one bright) of our system at $\Gamma$ (see Fig. \ref{sfig:2a}).
The incoming optical light ($E_{1u}$) interacts with the ground state $\ket{G}$ of the system (which is fully symmetric, $A_{1g}$) creating an excited state of symmetry $A_{1g}\otimes E_{1u}=E_{1u}$.
Therefore, the bright exciton corresponds to the $E_{1u}$ representation. 
The dark state is its Davydov partner, thus it must have opposite parity with respect to inversion symmetry\cite{multiBN} and it corresponds to the $E_{2g}$ representation, in analogy with the symmetry of the Davydov splitting for the acoustic phonons.

\begin{figure}[ht]
\includegraphics[width=0.8\columnwidth]{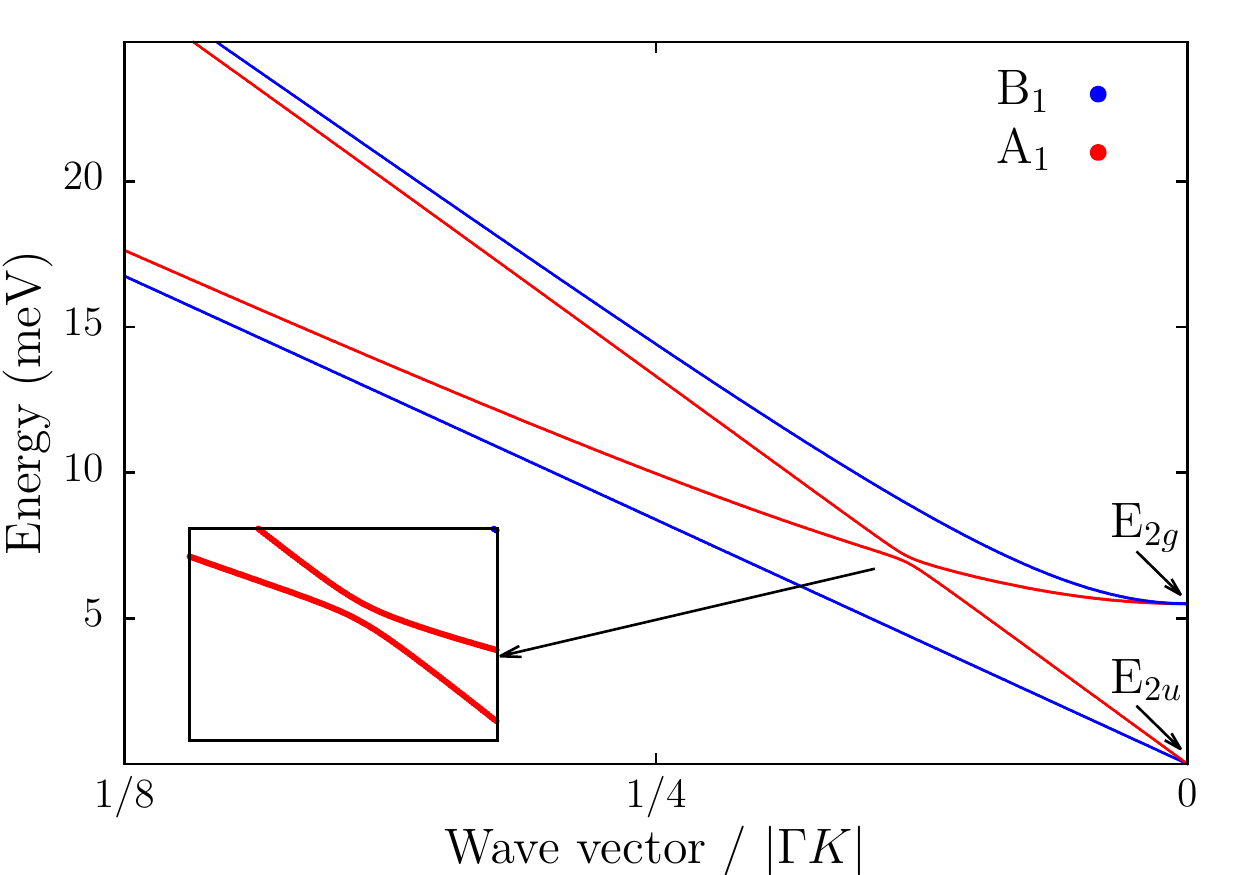}
\caption{Interplay between Davydov and symmetry splitting at very low wave vector for the transverse and longidutinal phonon modes (TA, TO$_1$, LA, LO$_1$ in the text). The dispersion of modes with symmetry A$_1$ (TO$_1$, LA) and B$_1$ (TA, LO$_1$) is shown in red and blue, respectively. \label{sfig:4d}}
\label{sfig:4b}
\end{figure}
Any irreducible representation of $D_{6h}$ will be a reducible representation of $C_{2v}$, and therefore can be expressed in terms of a linear combination of irreducible representations of $C_{2v}$.
These are the so-called compatibility relations that we will need to analyse indirect processes from $\Gamma$ to $\overline{q}$ (in particular, we want to describe the splitting of the $E_{1u}$ and $E_{2g}$ excitons). 
This applies also to the characters of the representations: if we define $\chi_{\mathcal{G}}(C_k)$ as the character of a (reducible) representation of group $\mathcal{G}$ with respect to symmetry operation $C_k$, the ``wonderful orthogonality theorem'' (according to Ref.~\onlinecite{Dresselhaus}) for characters establishes the coefficients $a_{\Gamma_i}$ of the linear combination associated with irreducible representation $\Gamma_i$. 
In our case, the formulas reduce to
\begin{equation*}
\begin{split}
\chi_{D_{6h}}(C_k)=\sum_{\Gamma_i} a_{\Gamma_i}\chi^{(\Gamma_i)}_{C_{2v}}(C_k) \\
a_{\Gamma_i} =\frac{1}{4}\sum_k \chi^{(\Gamma_i)}_{C_{2v}}(C_k) \chi_{D_{6h}}(C_k),
\end{split}
\end{equation*}
and we can compute the $a_{\Gamma_i}$ coefficients using the Tables \ref{t:D6h},  \ref{t:C2v} and \ref{t:conn}.
We find $A_{2u} \rightarrow B_2$ and both $E_{1u}$ and $E_{2g}$ splitting as $A_1+B_1$, confirming our previous identification of the dipole representations.
Therefore, the two indirect excitons labeled as $i1$ and $i2$ with momentum $\overline{q}$ can only have either $A_1$ or $B_1$ symmetry.

\subsection{Phonon symmetries and dispersion}
We list in Table \ref{t:phonons} the results of our DFPT calculations for the symmetries of the phonon modes at $\overline{q}$.
The formation of quasi-degenerate parallel phonon branches can be understood by zooming in on the region close to the $\Gamma$ point as in Fig. \ref{sfig:4d}. 
At $\Gamma$ we see the two degenerate TA and LA modes (E$_{2u}$ symmetry) at zero frequency, as well as their Davydov partner (TO$_1$ and LO$_1$ modes, E$_{2g}$ symmetry) $6$ meV above.
The large value of the splitting is due to the constructive interference of the Fourier components of the inter-layer interaction at zero wave vector.
When $q\neq 0$, the degenerate modes further split into two non-degenerate ones of symmetry A$_1$ and B$_1$. 
The two A$_1$ modes mix via an avoided crossing and then approach their respective B$_1$ mode. 
In this way two distinct Davydov pairs are formed, each one with a tiny energy splitting.
The low value of the splitting at finite $q$ is due to the destructive interference of the Fourier components of the inter-layer interaction.

\subsection{Selection rules for indirect absorption}
In order to analyse the indirect process, we first list in Table \ref{t:phonons} the results of our DFPT calculations for the symmetries of the phonon modes at $\overline{q}$.
Next we consider the time-dependent perturbation theory for a model excitonic Hamiltonian $H_0$ with exciton-radiation and exciton-lattice interactions as the perturbations: $H=H_0+V(t)=H_0+D\mathrm{e}^{-\mathrm{i}\omega t} +g_{\overline{q}} \mathrm{e}^{-\mathrm{i}\Omega_{\overline{q}} t} + h.c.$, where $\omega$ and $\Omega_{\overline{q}}$ are the photon and phonon frequencies, respectively, and $D$ and $g_{\overline{q}}$ the coupling operators (a sum over all phonon modes is assumed).
We want to qualitatively describe the phonon-assisted processes leading to the formation or annihilation of a finite-$q$ excitonic state $\ket{\psi_f}$.
We include only single-photon and single-phonon processes, and for simplicity we only consider phonon emission contributions. (Note that the following derivation is not meant to give a precise computational description of indirect absorption but that it serves only for the symmetry analysis of the involved phonons and excitons).
Then we can write the second-order Fermi golden rule expression for the transition probability per unit time as\cite{pastoriparr}
\begin{widetext}
\begin{equation}
\begin{split}
P^{II}_{f} = \frac{2\pi}{\hbar} \biggr\lvert \sum_{\alpha} \frac{\bra{\psi_f}g^{\dagger}_{\overline{q}}\ket{\psi_{\alpha}}\bra{\psi_{\alpha}}D\ket{G}}{E_{\alpha}-\hbar\omega} + \sum_{\alpha'} \frac{\bra{\psi_f}D\ket{\psi_{\alpha'}}\bra{\psi_{\alpha'}}g^{\dagger}_{\overline{q}}\ket{G}}{E_{\alpha'}-\hbar\Omega_{\overline{q}}} \biggr\rvert^2 \delta(E_f-\hbar\omega+\hbar\Omega_{\overline{q}}) + \\
   \frac{2\pi}{\hbar}\biggr\lvert \sum_{\alpha} \frac{\bra{\psi_f}g^{\dagger}_{\overline{q}}\ket{\psi_{\alpha}}\bra{\psi_{\alpha}}D^{\dagger}\ket{G}}{E_{\alpha}+\hbar\omega} + \sum_{\alpha'} \frac{\bra{\psi_f}D^{\dagger}\ket{\psi_{\alpha'}}\bra{\psi_{\alpha'}}g^{\dagger}_{\overline{q}}\ket{G}}{E_{\alpha'}-\hbar\Omega_{\overline{q}}} \biggr\rvert^2 \delta(E_f+\hbar\omega+\hbar\Omega_{\overline{q}}). 
\end{split}
\end{equation}
\end{widetext}
Here $\ket{\psi_{\alpha}}$ is an intermediate excitonic state with energy $E_{\alpha}$ and $\ket{G}$ is the ground state of the system.
The first term corresponds to the process of photon absorption with phonon emission creating the final excitonic state $\ket{\psi_f}$, while the second term describes the combined photon and phonon emission.
We notice immediately that if we take the static approximation for the transition probabilities (i.e. setting $\hbar\omega=\hbar\Omega_{\overline{q}}=0$ in the denominators) and $D=D^{\dagger}$ (e.g. the dipole operator), the squared quantities coincide for both photon absorption and emission.
This is the limit in which we performed our finite-difference calculations and is analogous to the Hall-Bardeen-Blatt theory of indirect absorption for independent particles.\cite{hbb}
The first term in the squared sum represents the creation of a direct virtual exciton $\ket{\psi_{\alpha}}$ by light as the intermediate step, followed by a scattering to the finite-$q$ state $\ket{\psi_f}$ via phonon emission.
The second term adds the contribution of the inverse process, when the virtual state $\ket{\psi_{\alpha'}}$ is created at finite-$q$ by a phonon, and then arrives at the energy of $\ket{\psi_f}$ by absorbing a photon.
\begin{figure}[ht]
\includegraphics[width=0.9\columnwidth]{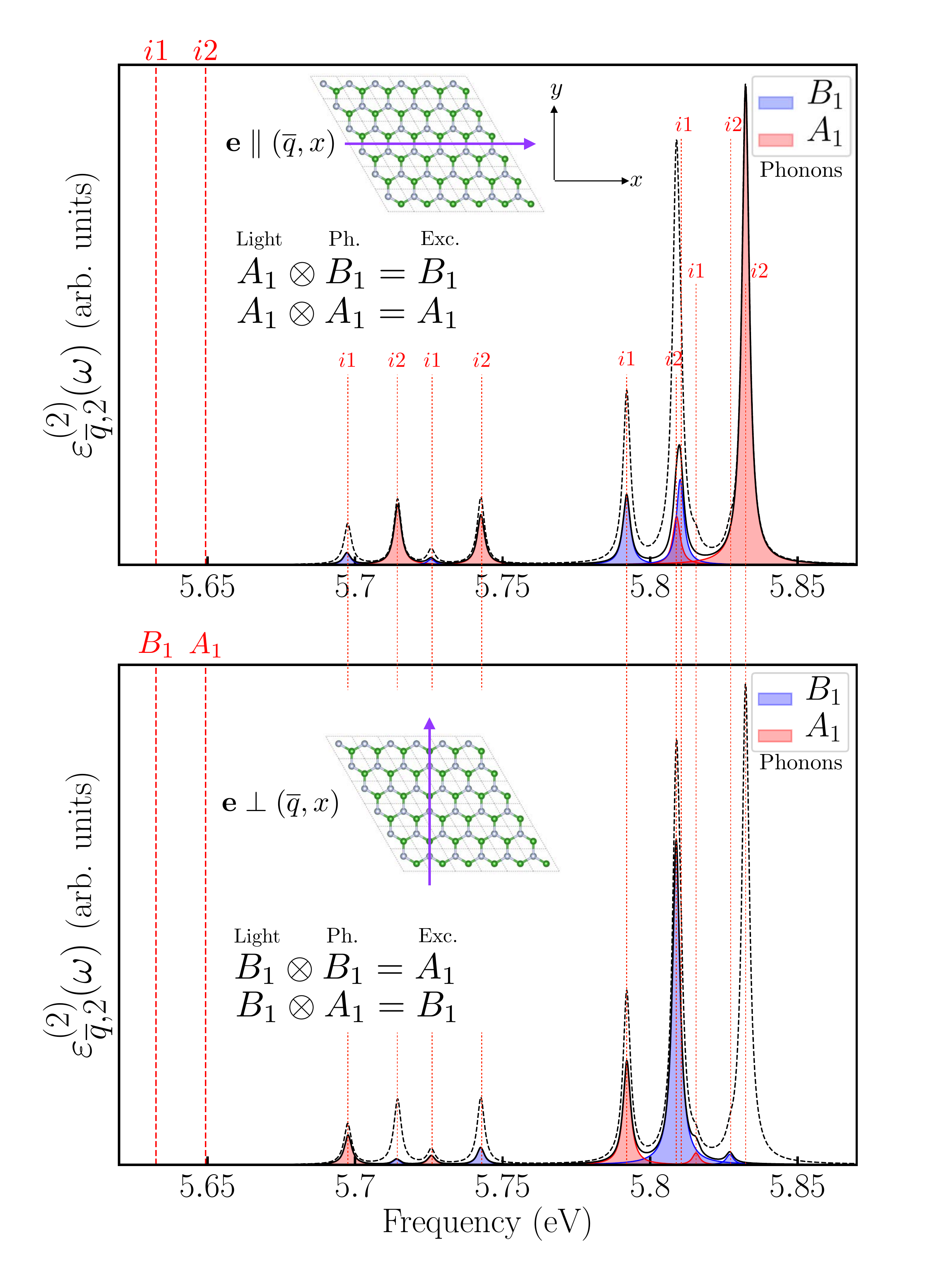}
\caption{\textit{Ab initio} results on exciton-phonon coupling, resolved by direction of light-polarization and symmetry of contributing phonon modes. The contribution to absorption, assisted by a phonon of wave vector $\bar{q}$ along the $x$-direction is plotted (solid black line) for light polarisation $\mathbf{e}$ along the $x$-direction (top frame) and the $y$-direction (bottom frame). The dashed black line represents the full result obtained by averaging over the two contributions. The red vertical lines serve as a guide for the eyes to understand which phonon-assisted peaks come from the coupling to excitons $i1$ or $i2$. The color of each phonon-assisted peak indicates the symmetry of the phonon mode responsible for it ($B_1$: blue, $A_1$: red).}
\label{sfig:4c}
\end{figure}
For the purposes of finding the selection rules, the two contributions are equivalent and thus we focus on the first one, $\bra{\psi_f}g^{\dagger}_{\overline{q}}\ket{\psi_{\alpha}}\bra{\psi_{\alpha}}D\ket{G}$.
The final allowed excitons in the energy window that we consider must have $A_1$ and $B_1$ symmetry and the first matrix element in the process, $\bra{\psi_{\alpha}}D\ket{G}$, imposes $E_{1u}$ as the only possible representation for the direct intermediate state $\ket{\psi_{\alpha}}$.
Since $E_{1u}\rightarrow A_1+B_1$ and $g^{\dagger}_{\overline{q}}$ transforms with the symmetry of the various phonon modes involved, for $g^{\dagger}_{\overline{q}}\ket{\psi_{\alpha}}$ we have the tensor product $(A_1+B_1+A_2+B_2)\otimes (A_1+B_1)$.
However, $A_1\otimes (A_2+B_2) = A_2+B_2$ and $B_1\otimes (A_2+B_2)=B_2+A_2$, therefore the phonon modes transforming as $A_2$ or $B_2$ cannot give the allowed final states and their coupling is forbidden.
We see from Table \ref{t:phonons} that these representations correspond to the out-of-plane Z phonon modes, while the in-plane ones T and L all transform as $A_1$ or $B_1$ and therefore are all allowed.
If we consider instead incoming light polarised out-of-plane ($A_{2u}\rightarrow B_2$), the picture changes and now $(A_1+B_1+A_2+B_2)\otimes B_2 = B_2+A_2+B_1+A_1$, meaning that if the polarisation is exclusively out-of-plane only the Z phonon modes can couple to excitons $i1$ and $i2$.

We find that our first-principles calculations respect the aforementioned selection rules remarkably well, as shown in Fig. \ref{sfig:4c}.
In particular, \textit{if light is polarised exclusively along the $x$/zigzag direction}, i.e. transforms as $A_1$ (top frame), then only the $TA$, $LO_1$, $TO_2$ and $LO_2$ phonon modes (all transforming as $B_1$, portrayed in blue) can couple to $i1$ forming phonon-assisted peaks, and only the $TO_1$, $LA$, $TO_3$ and $LO_3$ modes (all transforming as $A_1$, portrayed in red) can couple to $i2$.
Conversely, \textit{if light is polarised exclusively along the $y$/armchair direction}, i.e. transforms as $B_1$ (bottom frame), then only the $A_1$ phonon modes couple with $i1$, and only the $B_1$ modes couple with $i2$.
This unmistakably shows that exciton $i1$ (the one responsible for the luminescence spectrum) has $B_1$ symmetry, while exciton $i2$ transforms as $A_1$.
Additionally, we notice that the leading peak for absorption is due solely to the strong coupling between the $i2$ exciton and the $LO_3$ phonon mode: therefore, this peak completely disappears when light is polarised along the $y$ direction (that is, orthogonal to the $\overline{q}$ vector).

\begin{figure*}[ht]
\includegraphics[width=0.8\textwidth,trim={0cm, 6cm, 0cm, 0cm}]{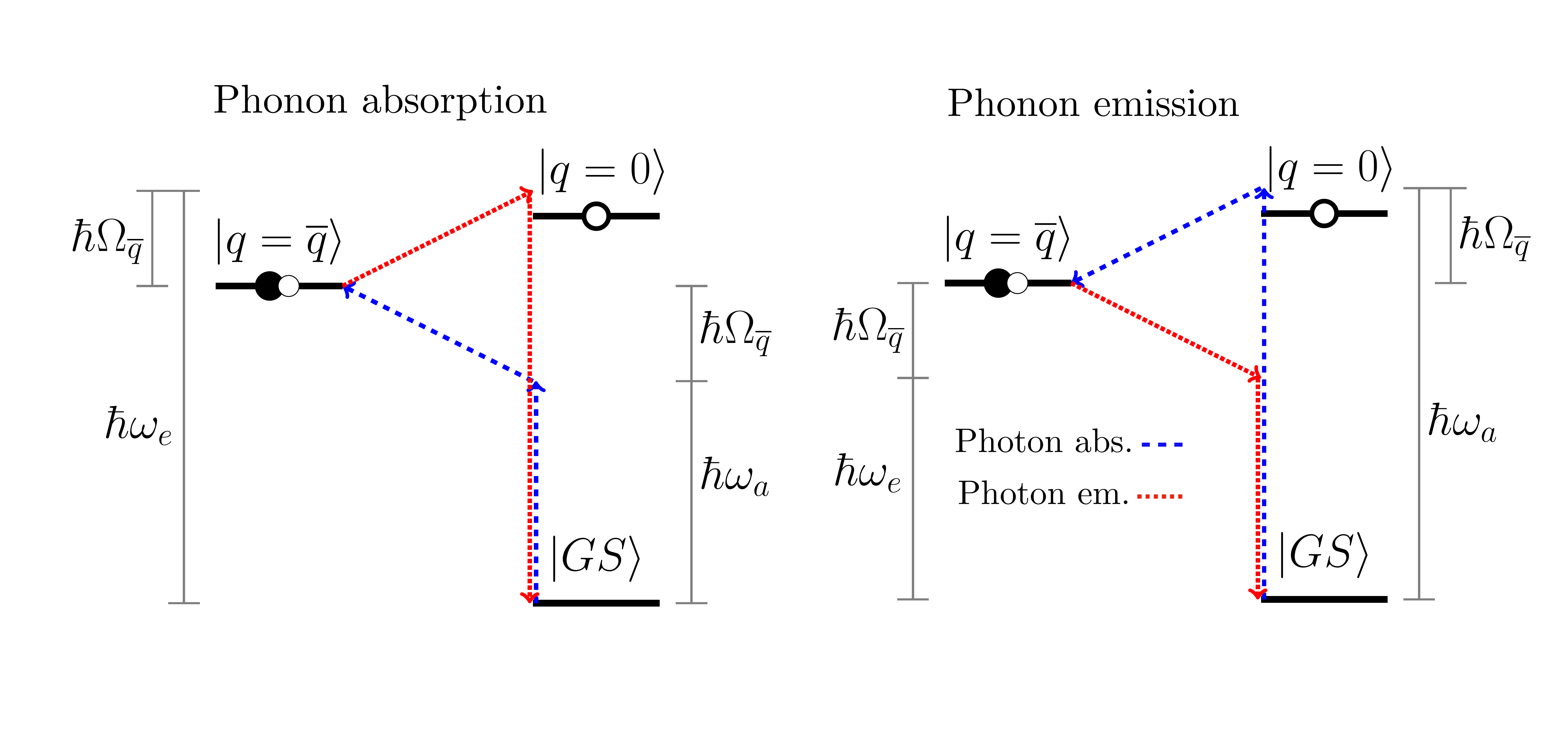}
\caption{Simplified two-exciton scheme for hBN, displaying the processes of photon absorption (dashed blue arrows) and emission (dotted red arrows) mediated by phonon absorption (left) and phonon emission (right). \label{sfig:5}}
\end{figure*}

\begin{figure}[ht!]
\includegraphics[width=0.9\columnwidth]{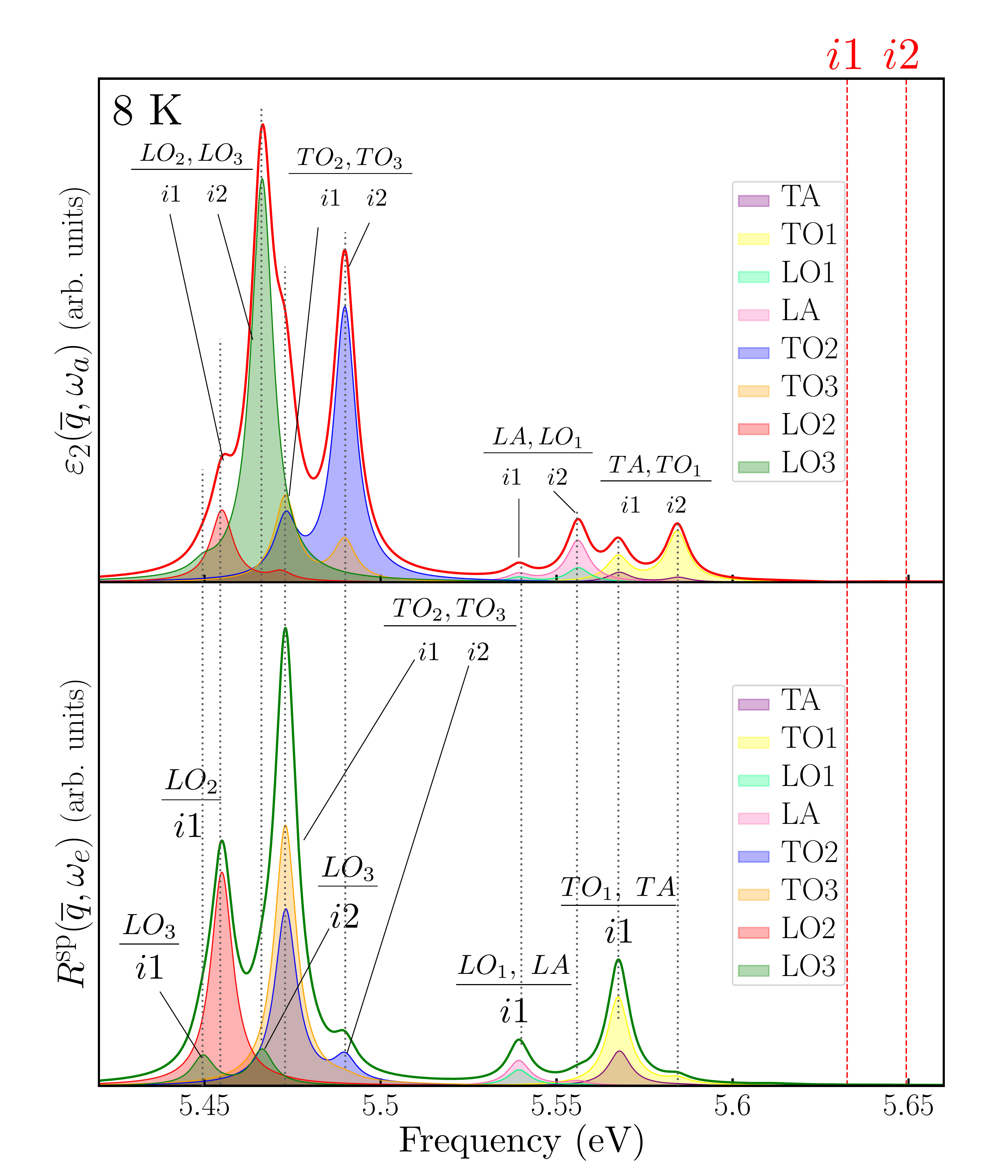}
\caption{Phonon-assisted emission in bulk hexagonal boron nitride. Spectral functions $\varepsilon_{2}(\overline{q},\omega_a)=\sum_{\lambda} \varepsilon_{2\lambda}(\overline{q},\omega_a)$ (red, top, $\mathrm{Im}\varepsilon^{em}_{\overline{q}}(\omega)$ in the main text) and $R^{\mathrm{sp}}(\overline{q},\omega_e)=\sum_{\lambda} R^{\mathrm{sp}}_{\lambda} (\overline{q},\omega_e)$ (green, bottom, $R^{\mathrm{sp}}_{\overline{q}}(\omega)$ in the main text) at $8$ K. The $\lambda$-components of the spectrum, belonging to the different phonon modes, are also plotted in various colours. The exciton-phonon couplings are labeled. The dotted vertical lines in correspondence with the peaks serve as a guide for the eyes. \label{sfig:5b}}
\end{figure}

\section{Van Roosbroeck-Shockley relation}
We give here a short account of the van Roosbroeck-Shockley relation,\cite{vRS} in order to add some context to its application to our results (Eq. $(4)$ of the main text).
We mainly adapt the contents of Refs. [\onlinecite{landsberg}] and [\onlinecite{bbwill}], where a more exhaustive treatment can be found.

\textbf{Direct transitions in the independent-particle case.}
Let us consider a steady state of photon absorption and emission processes between valence and conduction bands.
For simplicity of notation, the photon polarizations are not included.
The net absorption rate (per unit energy, per unit volume) will be given by the difference between the transition rates of absorption and stimulated emission processes:

\begin{widetext}
\begin{equation}\label{eq:abs}
R^\prime_{\mathrm{abs}} (\hbar \omega) = \mathcal{K}(\hbar\omega)\frac{2\pi}{\hbar} \frac{\mathcal{N}(\hbar \omega)}{N_k}\sum_{cvk} \mathcal{T}_{cvk} [ n^F_{vk} n^{F\prime}_{ck} - n^F_{ck} n^{F\prime}_{vk} ] \delta (\epsilon_{ck}-\epsilon_{vk} -\hbar \omega),
\end{equation}
\end{widetext}

where $\epsilon_{nk}$ is the energy of an electron in the $n$th band with wave vector $k$, $\mathcal{T}_{cvk}$ is the transition rate, $n^F_{nk}$ is the probability that state $\epsilon_{nk}$ is occupied with an electron, and $n^{F\prime}_{nk}$ the probability that it is occupied with a hole. 
We assume a time-independent Fermi-Dirac distribution for electronic occupations in the steady state, $n^F_{nk} = [1+\mathrm{e}^{(\epsilon_{nk}-\mu_e)/kT}]^{-1}$, with $\mu_e$ and $\mu_h$ being the chemical potentials for electrons and holes, respectively.
In the independent-particle case the transition rate can be taken as the optical matrix element in the dipole approximation: $\mathcal{T}_{cvk} = |\bra{ck}\hat{D}\ket{vk}|^2$. 
We can similarly write the expression for the spontaneous emission rate,

\begin{equation}\label{eq:em}
R^{sp}(\hbar\omega) =\mathcal{K}(\hbar\omega) \frac{2\pi}{\hbar}\frac{\mathcal{G}(\hbar\omega)}{N_k}\sum_{cvk} \mathcal{T}_{cvk}n^F_{ck} n^{F\prime}_{vk}  \delta (\epsilon_{ck}-\epsilon_{vk} -\hbar \omega).
\end{equation}

The spontaneous emission is only proportional to the photon density of states, $\mathcal{G}(\hbar\omega)$, while both absorption and stimulated emission are proportional to the the photon density per unit energy, $\mathcal{N}(\hbar\omega)$. 
If we define an average photon number, $\overline{\mathcal{N}}$, these two quantities are related by the total photon density $\int \mathcal{N}(\hbar\omega)\mathrm{d}\hbar\omega = \int \overline{\mathcal{N}} \mathcal{G}(\hbar\omega) \mathrm{d}\hbar\omega$.
The dimensional term $\mathcal{K}(\hbar\omega)$ is made of quantities mainly related to the the electro-magnetic field.
It is important to notice that this term is also frequency-dependent.
We now list the expressions for the optical quantities involved.
\begin{equation*}
\begin{split}
& \mathrm{Dimensional \ factor} \quad \mathcal{K}(\hbar\omega)=\frac{2\pi e^2 \hbar^2}{m^2 V} \frac{1}{n_r(\hbar\omega)^2 \hbar\omega}        \\
& \mathrm{Photon \ density \ of \ states} \quad \mathcal{G}(\hbar\omega)=\frac{1}{\pi^2 c^2 \hbar^3}\frac{n_r(\hbar\omega)^2 (\hbar\omega)^2}{V_g(\hbar\omega)}\\
& \mathrm{Group \ velocity} \quad V_g(\hbar\omega)=\frac{c}{n_r(\hbar\omega)+\omega \frac{\partial n_r(\hbar\omega)}{\partial\omega}}\\
& \mathrm{Velocity \ of \ energy \ transfer} \quad V_{en}(\hbar\omega)=\frac{c}{n_r(\hbar\omega)}\\
& \mathrm{Incoming \ photon \ flux} \quad \mathcal{F}(\hbar\omega)=\mathcal{N}(\hbar\omega)V_{en}(\hbar\omega)
\end{split}
\end{equation*} 

The absorption coefficient $\alpha(\hbar\omega)$ can be written in terms of the absorption rate as $R^\prime_{\mathrm{abs}} (\hbar \omega) = \mathcal{F}(\hbar \omega) \alpha(\hbar\omega)$.
Finally, we observe that independently from the specific $(cvk)$ transition considered, the following relation always holds:
\begin{equation*}
\frac{n^F_{ck} n^{F\prime}_{vk}}{ n^F_{vk} n^{F\prime}_{ck} - n^F_{ck} n^{F\prime}_{vk}} = \frac{1}{\mathrm{e}^{(\hbar\omega-(\mu_e-\mu_h))/kT}-1}\approx \mathrm{e}^{-(\hbar\omega-\Delta\mu)/kT}
\end{equation*}
Then, by putting everything together and comparing Eqs. \eqref{eq:abs} and \eqref{eq:em}, we find $R_{\mathrm{sp}}(\hbar\omega)=\mathcal{G}(\hbar\omega)V_{en}(\hbar\omega) \alpha(\hbar\omega) \mathrm{e}^{-(\hbar\omega-\Delta\mu)/kT}$. 
The prefactor reads
\begin{equation*}
\begin{split}
\mathcal{G}(\hbar\omega)V_{en}(\hbar\omega)&=\frac{n_r(\hbar\omega)^2 (\hbar\omega)^2}{\pi^2 c^2 \hbar^3}\frac{V_{en}(\hbar\omega)}{V_g(\hbar\omega)}\\
&=\frac{n_r(\hbar\omega)^2 (\hbar\omega)^2}{\pi^2 c^2 \hbar^3}\left[1+\omega\frac{\partial \mathrm{ln} n_r(\hbar\omega)}{\partial \omega} \right]
\end{split}
\end{equation*}
In dielectric mediums usually we have $V_{en}(\hbar\omega)=V_g(\hbar\omega)$, which corresponds to neglecting the frequency-dependent term in the square brackets.
This leads us to the van Roosbroeck-Shockley relation:

\begin{equation}\label{eq:vRS}
\begin{split}
R_{\mathrm{sp}}(\hbar\omega)=\frac{n_r(\hbar\omega)^2(\hbar\omega)^2}{\pi^2 c^2 \hbar^3} \alpha(\hbar\omega) \mathrm{e}^{-(\hbar\omega-\Delta\mu)/kT} \\ = \frac{n_r(\hbar\omega)(\hbar\omega)^3}{\pi^2 c^3 \hbar^4} \varepsilon_2(\hbar\omega) \mathrm{e}^{-(\hbar\omega-\Delta\mu)/kT},
\end{split}
\end{equation}
where for the last equality we have used $\alpha(\hbar\omega)=\hbar\omega \varepsilon_2(\hbar\omega)/(n_r(\hbar\omega)\hbar c)$.

\textbf{Direct transitions in the exciton case.} 
If the absorption/emission features are dominated by the creation/annihilation of electron-hole bound pairs, it is sufficient to replace the Bose-Einstein/Boltzmann factor in Eq. \eqref{eq:vRS} with one more appropriate to describe the occupation of excitonic states.
In the main text we use the distribution $N_B(\hbar \omega) = \mathrm{e}^{-(\hbar \omega - \mu^*)/kT}$, with $\mu^*$ fixed to the energy of the lowest-bound exciton.
Now $\alpha(\hbar\omega)$ or equivalently $\varepsilon_2 (\hbar \omega)$ are computed including excitonic effects (we can always obtain the full refractive index as $n_r(\hbar \omega) = \sqrt{(\sqrt{\varepsilon_1(\hbar\omega)^2+\varepsilon_2(\hbar\omega)^2} + \varepsilon_1(\hbar\omega))/2}$).
This means that for the transition rate we will consider the excitonic oscillator strengths, $\mathcal{T}_S = \lvert \sum_{cvk} \Phi^S_{cvk}\bra{ck}\hat{p}\ket{vk}\rvert^2$, with the external sum now running over the exciton index $S$.

\textbf{Indirect transitions.} 
In this case we have to take into account that the energy of a photon absorbed ($\hbar\omega_a$) and that of a photon emitted ($\hbar\omega_e$) in a process mediated by the same phonon are not the same, and they are both different from the energy of the electronic transition ($\hbar\omega$).
In particular, with the help of Fig. \ref{sfig:5}, we can write the following relations:
\begin{equation*}
\begin{split}
& \hbar\omega_e = \hbar\omega_a \pm 2\hbar \Omega_\lambda \\ 
& \hbar\omega_e = \hbar\omega \pm \hbar\Omega_\lambda \\ 
& \hbar\omega_a = \hbar\omega \mp \hbar\Omega_\lambda,
\end{split}
\end{equation*}
where $\Omega_\lambda$ is the frequency of the phonon assisting the transition and the upper and lower signs refer to the cases of phonon absorption and emission, respectively.
We need to write a generalised form of the van Roosbroeck-Shockley relation that takes these energy differences into account.
Focusing on the case of an indirect transition mediated by the emission of a single phonon of branch $\lambda$ and momentum $q$, the second-order absorption and emission rates (meaning per unit time, energy and volume) can be expressed as (back in the independent-particle case)
\begin{widetext}
\begin{equation}
\begin{split}
& R^\prime_{\mathrm{abs}} (\hbar \omega_a) = \mathcal{K}(\hbar\omega_a)\frac{2\pi}{\hbar} \frac{\mathcal{N}(\hbar \omega_a)}{N_k}\sum_{cvk} \mathcal{T}^{(2)}_{cvkq\lambda} (n_{q\lambda}+1)[ n^F_{vk} n^{F\prime}_{ck+q} - n^F_{ck+q} n^{F\prime}_{vk} ] \delta (\epsilon_{ck+q}-\epsilon_{vk} +\hbar\Omega_{q\lambda} - \hbar \omega_a) \\
& R^{sp}(\hbar\omega_e) =\mathcal{K}(\hbar\omega_e)\frac{2\pi}{\hbar}\frac{\mathcal{G}(\hbar\omega_e)}{N_k}\sum_{cvk} \mathcal{T}^{(2)}_{cvkq\lambda}(n_{q\lambda}+1) n^F_{ck+q} n^{F\prime}_{vk}  \delta (\epsilon_{ck+q}-\epsilon_{vk} -\hbar\Omega_{q\lambda} -\hbar \omega_e),
\end{split}
\end{equation}
\end{widetext}
where $n_{q\lambda}$ represents the Bose-Einstein distribution for phonons (we use $n_B$ in the main text).

Then, considering $\alpha (\hbar\omega_a) = R^\prime_{\mathrm{abs}} (\hbar \omega_a)/\mathcal{F}(\hbar\omega_a)$ and writing the frequency-dependent functions explicitly, we can write the final results:
\begin{equation}\label{eq:generalisedRS}
\begin{split}
R^{sp}(\hbar\omega_e) &= \frac{n_r (\hbar\omega_e) n_r (\hbar\omega_a) (\hbar\omega_e) (\hbar\omega_a)}{\pi^2 c^2 \hbar^3} \alpha(\hbar\omega_a)  \mathrm{e}^{-(\hbar\omega-\Delta\mu)/kT}  \\ &= \frac{n_r(\hbar\omega_e) (\hbar\omega_e) (\hbar\omega_a)^2 }{\pi^2 c^3 \hbar^4} \varepsilon_2(\hbar\omega_a)  \mathrm{e}^{-(\hbar\omega-\Delta\mu)/kT}.
\end{split}
\end{equation}

Now we just need to replace $\hbar\omega_a$ with $\hbar\omega_e - 2\hbar\Omega_{q\lambda}$ and use the Boltzmann distribution for excitons.
The emission spectra shown in Fig. 3 and 4 of the main text are obtained by using Eq. \eqref{eq:generalisedRS} (Eq. (4) in the main text) and summing over all phonon modes with momentum $q=\overline{q}$ corresponding to the indirect gap.
A comparison between $\varepsilon_2(\overline{q},\hbar\omega_a)$ (where all phonon-assisted peaks are mirrored with respect to the energy of the excitonic state involved) and $R^{sp}(\hbar\omega_e)$, the full RS relation, is shown in Fig. \ref{sfig:5b}.

\FloatBarrier

\bibliography{excphon}